\documentclass[11pt]{article}
\usepackage{cite,amsmath,amsfonts,amsthm,fullpage}
\usepackage{youngtab}
\usepackage{cite,amsmath,amsfonts,amsthm,fullpage, amssymb, mathtools, url}
\usepackage[usenames,svgnames]{xcolor}
\usepackage{youngtab, enumitem, comment, mathdots, graphicx, float}
\usepackage[symbol]{footmisc}
\usepackage[pagebackref=false]{hyperref}

\theoremstyle{plain}
\theoremstyle{definition}
\newtheorem{theorem}{Theorem}[section]
\newtheorem{lemma}[theorem]{Lemma}
\newtheorem{definition-theorem}[theorem]{Definition-Theorem}
\newtheorem{definition-proposition}[theorem]{Definition-Proposition}

\newtheorem{proposition}[theorem]{Proposition}
\newtheorem{corollary}[theorem]{Corollary}
\newtheorem{example}{Example}[section]
\newtheorem{examples}{Example}[subsection]

\newtheorem{remark}{Remark}[section]
\newtheorem{remarks}{Remarks}[section]

\numberwithin{equation}{section} 

\DeclareMathOperator{\ch}{ch}

\def\Tr{\mathrm {Tr}}

\def\det{\mathrm {det}}

\def\Pf{\mathrm {Pf}}

\def\res{\mathop{\mathrm {res}}\limits}
\def\nn{\nonumber}

\def\res{\mathop{\mathrm{res}}\limits}

\def\&{&{\hskip -20pt}}

\def\be{\begin{eqnarray}}
\def\ee{\end{eqnarray}}

\def\bt{\begin{theorem}}
\def\et{\end{theorem}}

\def\bp{\begin{proposition}}
\def\ep{\end{proposition}}

\def\bex{\begin{example}\small \rm}
\def\eex{\end{example}}

\def\bexs{\begin{examples}\small \rm}
\def\eexs{\end{examples}}

\def\br{\begin{remark}\small \rm}
\def\er{\end{remark}}

\def\Mb{\mathbf{M}}
\def\Nb{\mathbf{N}}

\def\Zb{\mathbf{Z}}

\def\pb{\mathbf{p}}

\def\tb{\mathbf{t}}

\def\xb{\mathbf{x}}

\newcommand{\DP}{\mathop\mathrm{DP}\nolimits}
\newcommand{\OP}{\mathop\mathrm{OP}\nolimits}

\newcommand{\f}{\textbf{f}}

\def\bp{\begin{Proposition}\rm}
\def\ep{\end{Proposition}}
\def\bc{\begin{corollary}}
\def\ec{\end{corollary}}
\def\bl{\begin{lemma}\em}
\def\el{\end{lemma}}
\def\br{\begin{remark}\rm\small}
\def\er{\end{remark}}
\def\brs{\begin{remarks}.\\ \rm\
\begin{enumerate}}
\def\ers{\end{enumerate}\end{remarks}}
\def\bea{\begin{eqnarray}}
\def\eea{\end{eqnarray}}

\begin{document}

\begin{center}
\begin{small}
FIAN/TD-07/20 \hfill\phantom.\\
IITP/TH-19/20 \hfill\phantom.\\
ITEP/TH-30/20 \hfill\phantom.\\
MIPT/TH-17/20 \hfill\phantom.\\
\end{small}

\vspace{-0.9cm}

{\bf
\hfill to the memory of\\
\hfill  Boris Dubrovin\\
}
\end{center}

\vspace{1cm}

\begin{center}

\begin{Large}\fontfamily{cmss}
\fontsize{17pt}{27pt}
\selectfont
	\textbf{Around spin Hurwitz numbers }
	\end{Large}
	
\bigskip \bigskip
\begin{large}A. D. Mironov$^{a,b,c}$\footnote{mironov@lpi.ru; mironov@itep.ru},
A. Morozov$^{d,b,c}$\footnote{morozov@itep.ru},
\fbox{S. M. Natanzon$^{e,b}$}\footnote{Sergey Natanzon passed away on December, 7, 2020},
A. Yu. Orlov$^{f}$\footnote{orlovs55@mail.ru}
 \end{large}
 \\
\bigskip

\begin{small}
$^a$ {\it Lebedev Physics Institute, Moscow 119991, Russia}\\
$^b$ {\it ITEP, Moscow 117218, Russia}\\
$^c$ {\it Institute for Information Transmission Problems, Moscow 127994, Russia}\\
$^d$ {\it MIPT, Dolgoprudny, 141701, Russia}\\
$^e$ {\it HSE University, Moscow, Russian Federation}\\
$^f${\em Shirshov Institute of Oceanology, RAS, Nahimovskii Prospekt 36, Moscow 117997, Russia }
\end{small}
 \end{center}
\medskip

\begin{abstract}
We present a review of the spin Hurwitz numbers, which count the ramified coverings with spin structures.
They are related to peculiar $Q$ Schur functions, which are actually related to characters of the Sergeev group. This allows one to put the whole story into the modern context of matrix models and integrable hierarchies. Hurwitz partition functions are actually broader than the conventional $\tau$-functions, but reduce to them in particular circumstances. We explain how special $d$-soliton $\tau$-functions of the KdV and Veselov-Novikov hierarchies
generate the spin Hurwitz numbers $H^\pm\left( \Gamma^b_d \right)$
and $H^\pm\left( \Gamma^b_d,\Delta \right)$. The generating functions of the spin Hurwitz numbers are
hypergeometric $\tau$-functions of the BKP integrable hierarchy, and we present their fermionic realization.
We also explain how one can construct $\tau$-functions of this type entirely in terms of the $Q$ Schur functions.
An important role in this approach is played by factorization formulas for the $Q$ Schur functions on special loci.
\end{abstract}

\bigskip

Boris Dubrovin was one of the brightest minds in modern mathematics,
in fact, his well-known achievements cover only a small part of all
what he thought about and planned to do.
Long ago, as young students, we were inspired by the brilliant volume
\cite{SovGeom}, to which Boris also contributed a big part of his knowledge and vision.
During his entire life in science, he tried to make new fields as clear and transparent,
as happened to the classical subjects in that unique textbook.

This paper was supposed to have a single author, Sergey Natanzon,
and was planned as his personal tribute to his close friend, Boris.
Unfortunately, Sergey was not given a chance to fulfill this duty.
The present note is a brief review of what he taught us about the
the increasingly important subject of {\it spin} Hurwitz numbers,
and it certainly lacks the vision which Sergey had on it and
which we, survivors,  will still need to rediscover
in our future work.

\newpage

\tableofcontents

\section{Introduction \label{introduction}}

Hurwitz numbers \cite{Hur,Fro,Burn,Di} count ramified coverings, and their significance in physics
is dictated by the role that complex curves play in string theory.
Applications range from the free field calculus on ramified Riemann surfaces
\cite{Kni} to genus expansion in matrix models and topological recursion \cite{AMM/EO,AMM}.
The spin Hurwitz numbers \cite{EOP,G} do just the same things to Riemann surfaces with
spin structures labeled by theta-characteristics.
The matrix model counterparts in this case appear to be the cubic Kontsevich \cite{Ko,GKM}
and Brezin-Gross-Witten (BGW) \cite{BGW,AMM} models,
which are non-obvious generalizations of the Hermitian matrix model \cite{UFN3}
equivalent to a quadratic counterpart of Kontsevich theory \cite{Che,versus}.
As usual, the most efficient technique to develop spin Hurwitz calculus
relies on algebraic approach, which finally puts the problem into
the framework of integrable systems.
The basic of this approach includes the following ingredients \cite{MMN2019}:

\begin{itemize}

\item{A relevant substitute of the Schur functions, which in the case
of cubic Kontsevich and BGW models is provided by the $Q$ Schur functions}

\item{Their relation to the Hall-Littlewood polynomials \cite{Mac}}

\item{Their relation to characters of the Sergeev algebra \cite{Serg},
which provides a relevant generalization of symmetric group characters}

\item{The Fr\"obenius formula, expressing the  spin Hurwitz numbers through
the Sergeev characters and the $Q$ Schur functions}

\item{Commuting system of cut-and-join $\hat W$ operators \cite{MMN12,MMN2019},
which have $Q$ Schur functions as their common eigenfunctions
and the Sergeev characters as eigenvalues}

\item{Free fermion representations \cite{DJKM1,JM,O2003,You}, which allow one to represent
$Q$ Schur functions as Pfaffians}

\item{Integrability properties of the spin Hurwitz numbers.
As usual \cite{MMN,MMN12,MMS}, Hurwitz $\tau$-functions form a broader and
still uncomprehensible variety, but, in special cases,
we get soliton and other solutions to the BKP hierarchy \cite{DJKM1,JM}}

\item{Hypergeometric $\tau$-functions of the BKP hierarchy \cite{O2003,NimmoOr}.
By definition, they are bilinear in the $Q$ Schur functions
with coefficients of a very special product form ($\DP$ denotes the strict partitions)}
\be\label{1}
\tau_{BKP}\{p_k,p_k^*\} = \sum_{\alpha\in\DP} Q_\alpha\{p_k\}Q_\alpha\{p_k^*\} \cdot \prod_{i=1}^{\ell(\alpha)} f(\alpha_i)
\ee
Important examples are provided by the ratios
\be
\frac{Q_{N\alpha}\{\delta_{k,r}\}}{Q_\alpha\{\delta_{k,r}\}} = \prod_{i=1}^{\ell(\alpha)} f(\alpha_i)
\ \ \ \ {\forall} \ {\rm coprime} \  N,r
\ee
where $\ell(\alpha)$ denotes the number of lines in the Young diagram $\alpha$, and $N\alpha$ denotes the Young diagram with lengths $N\alpha_i$.
In particular, for the cubic Kontsevich model \cite{MMq},
\be\label{Kon}
\tau_{K_3}\{p_k\} = \sum_{\alpha\in\DP} Q_\alpha\{p_k\}Q_\alpha\{\delta_{k,3}\}\cdot
\frac{Q_{\alpha}\{\delta_{k,1}\}}{Q_{2\alpha}\{\delta_{k,1}\}}
\frac{Q_{2\alpha}\{\delta_{k,3}\}}{Q_\alpha\{\delta_{k,3}\}}\cdot{1\over 2^{\ell(\alpha)}}
\ee
while, for the BGW model \cite{Alex},
\be\label{BGW}
\tau_{BGW}\{p_k\} = \sum_{\alpha\in\DP} Q_\alpha\{p_k\}Q_\alpha\{\delta_{k,1}\}\cdot
\left(\frac{Q_{\alpha}\{\delta_{k,1}\}}{Q_{2\alpha}\{\delta_{k,1}\}}\right)^2\cdot{1\over 2^{\ell(\alpha)}}
\ee

\end{itemize}

We refer to the very recent paper \cite{Alex}
for some complimentary details.

As to the ordinary integrability, its intimate relation to characters is well known.
The Schur functions are themselves solutions to the Hirota bilinear equations,
and general KP/Toda $\tau$-functions are their linear combinations
with the coefficients satisfying the Pl\"ucker relations,
which have determinants and their free fermion realizations
as natural solutions.
The $Q$ Schur functions play the same role for the BKP hierarchies,
only solutions are now Pfaffians, which can be described in terms of
``neutral" fermions.
These simple facts are, however, not fully trivial: for example, the
KdV hierarchy, which can be considered as a reduction of the both
KP and BKP hierarchies \cite{DJKM1,JM,DJKM2,DJKM,Alex2},
possesses as solutions only the very special $Q$ Schur functions
$Q_{[1]}$, $Q_{[2,1]}$, $Q_{3,2,1}$, $\ldots$\,,
$Q_{[\ldots,4,3,2,1]}\sim \hbox{Schur}_{[\ldots,4,3,2,1]}$,
which are the only Schur functions independent of even time-variables $p_{2k}$ and proportional to the corresponding $Q$ Schur functions. All other $Q$ Schur functions, which are all independent of $p_{2k}$,
do not solve the KdV hierarchy.

In the present paper, we review some auxiliary aspects,
related to the free-fermion description of the $Q$ Schur functions, the
BKP and KdV hierarchies.
In particular, as in the case of ordinary Hurwitz numbers,
the lowest ``cut-and-join" $\hat W$ operators commute with the
BKP Hirota equations and generates an especially simple hypergeometric
``tau"-function,
which can be also considered as an infinite-soliton
$\tau$-function of the KdV hierarchy.
Technically, the BKP $\tau$-function can be defined as
a direct counterpart of the Toda lattice $\tau$-function, but for the matrix
elements of $SL(\infty)$ generated by a restricted set
of ``neutral" fermions and depending only on odd sets of times.
Proper weights made of exponentials of power sums of Young diagram lengths in (\ref{1}) (``completed cycles")
in this formalism provides the hypergeometric $\tau$-functions,
which are easy and
very straightforward to work with.
The full (spin) Hurwitz $\tau$-functions involve far more complicated
weights made out of all (Sergeev)
symmetric characters, and they provide an important
generalization beyond (B)KP theory,
which still awaits an efficient language and deep investigation.

\paragraph{Notation.}
Throughout the paper, we denote through $[x]$ the integer part of a number, through $\{x\}$ its fractional part. For an integer $k$, $(k)_r=r\{k/r\}$ denotes the value mod $r$. For the strict partition $\alpha$, $\ell(\alpha)$ is the number of parts, and $\bar\ell(\alpha):=2\cdot\left[{\ell(\alpha)+1\over 2}\right]$.

\section{Hurwitz numbers}

\subsection{Classical Hurwitz numbers}

Consider a compact Riemann surface $S$ of genus $g$ with a finite number of points $x_1,\dots,x_n\in S$. Consider a set of Young diagrams $(\Delta^1,\dots,\Delta^n)$ of the same degree $d=|\Delta_i|$. The lengths of the rows $\Delta^i_{1},\dots,\Delta^i_{\ell_i}$ of the Young diagram  $\Delta^i$  give the partition of the number $d$.

Denote by $\widetilde{M}(\Delta^1,\dots,\Delta^n)$  the set of holomorphic mappings of compact Riemann surfaces $\varphi: P\rightarrow S$, whose critical values lie in $\{x^1,\dots,x^n\}$, and the pre-images $\varphi^{-1}(x^i)$ consist of points, where $\varphi$ has degrees  $\Delta^i_{1},\dots,\Delta^i_{\ell_i}$. We call the mappings $\varphi: P\rightarrow S$  and $\varphi': P'\rightarrow S$ as equivalent if there exists a biholomorphic mapping $\phi$ such that $\varphi = \varphi' \phi $. Let $ M(\Delta^1, \dots, \Delta^n) $ denote the set of equivalence classes in the set $ \widetilde{M} (\Delta^1, \dots, \Delta^n)$.

The classical Hurwitz number \cite{Di} is the number
\be
H_d(g|\Delta^1,\dots,\Delta^n)=\sum\limits_{\varphi\in M(\Delta^1,\dots,\Delta^n)}\frac{1}{|Aut(\varphi)|}
\ee
There is the Fr\"obenius formula that gives a combinatorial expression for the Hurwitz numbers \cite{Burn},
\be\label{Fro1}
H_d(g|\Delta^1,\dots,\Delta^n)=\frac{[\Delta_1]\dots[\Delta_k]}{(d!)^2} \sum\limits_{R}\frac{\psi_R(\Delta_1)\dots\psi_R(\Delta_k)}
{\psi_R(1)^{(k-2)}}
\ee
where $[\Delta]$ is the number of permutations of the cyclic type $\Delta$,  i.e. the number of elements in the conjugacy class of the symmetric group $\mathfrak{S}_d$ given by the Young diagram $\Delta$, $|\Delta|=d$; $\psi_R(\Delta)$ is value of the character $\psi_R$ of the representation $R$ of the symmetric group $\mathfrak{S}_d$ on the permutation of cyclic type $\Delta$, $\psi_R(1)$ is the value on the permutation with all unit cycles, $\Delta=[\underbrace{1,\ldots,1}_{d\ \rm{times}}]$,
and the sum is taken over all characters of irreducible representations of $\mathfrak{S}_d$.

\vspace{2ex}

Among the classical Hurwitz numbers, we will be interested only in the so-called double Hurwitz numbers
\be
H_d(\Delta,\Delta^*,b)=H_d(0|\Delta,\Delta^*,\Gamma_2^b)
\ee
Here $\Gamma_2$ is a Young diagram with one row of length 2 and the rest of rows of length 1. The $\Gamma_2^b$ means a set from $b$ diagrams $\Gamma_2$.

Consider two infinite sets of variables $p=(p_1,p_2,\dots)$ and $p^*=(p_1^*,p_2^*,\dots)$. Associate monomial $p_{\Delta}=p_{\Delta_1}\cdots p_{\Delta_{\ell}}$ to the Young diagram $\Delta=[\Delta_{1},\dots,\Delta_{\ell}]$  and monomial
$p_{\Delta^*}^*=p_{\Delta_1^*}^*\cdots p_{\Delta_{\ell}^*}^*$ to Young diagram $\Delta^*=[\Delta_{1}^*,\dots,\Delta_{\ell}^*]$.
To the double Hurwitz number $H_d(\Delta,\Delta^*,b)$ we associate the monomial
$H_d(\Delta,\Delta^*,b)p_{\Delta}p_{\Delta^*}^*$\ .
As a generating function for the double Hurwitz numbers, one usually considers the function proposed in \cite{Ok}
\be
\tau(\Delta,\Delta^*,\beta,q)=\sum\limits_{d>0}\sum\limits_{|\Delta|=|\Delta^*|=0}
\sum\limits_{b,\Delta,\Delta^*}q^d\frac{\beta^b}{b!}H_d(\Delta,\Delta^*,b)p_{\Delta}p_{\Delta^*}^*
\ee
According to \cite{GKM2},\cite{OS},\cite{Ok},\cite{AMMN1} this function is a $\tau$-function of the 2D Toda lattice (it was called hypergeometric in \cite{OS}).

Moreover, according to \cite{GJW, MMN12}, it satisfies the \textit{cut-and-join} equation
\be
\frac{\partial\tau(\Delta,\Delta^*,\beta,q)}{\partial\beta}= W \tau(\Delta,\Delta^*,\beta,q)
\ee
where
\be
W=\frac{1}{2}\sum\limits_{a,b>0}\Big( (a+b)p_ap_b\frac{\partial}{\partial p_{a+b}}
+ ab p_{a+b}\frac{\partial^2}{\partial p_{a}\partial p_{b}}\Big)
\ee

\vspace{2ex}

\subsection{Spin Hurwitz numbers}

A line bundle $L$ on a Riemann surface is called a spin bundle if the tensor square of $L$
is isomorphic to the cotangent bundle. The parity of the space of holomorphic sections of the bundle $L$ is called the parity of the bundle (see \cite{At,Mu}) and is denoted by $Arf(L)\in \{0,1\}$. The surface of genus $0$ has exactly one spin bundle, and it is even.

Consider a holomorphic mapping $\varphi: P \rightarrow S$ whose critical point orders are odd. Such a mapping associates the spin bundle $L$ onto $S$ with the spin bundle $ \varphi^*(L) $ onto $P$. Its parity $ Arf (\varphi) = Arf(\varphi^*(L)) $ depends only on $ \varphi $ and $Arf(L) $.

The spin Hurwitz number was defined in \cite{EOP} as
\be
H_d^{{\rm Arf}(L)}(g|\Delta_1,\dots,\Delta_n)=\sum\limits_{\varphi\in M(\Delta_1,\dots,\Delta_n)}
\frac{(-1)^{{\rm Arf}(\varphi)}}{|{\rm Aut}(\varphi)|}
\ee
Depending on whether the parity of the bundle is even or odd, later on, we use the superscripts $+$ and $-$ respectively.

We are only interested in the double spin Hurwitz numbers, which, in this case, have the form
\be
H_d^{sp} (\Delta, \Delta^*, b) = H_d^0 (0 | \Delta, \Delta^*, \Gamma_3^b)
\ee
where $ \Gamma_3 $ is a Young diagram with one row of length 3 and other rows of length 1, and $\Gamma_3^b $ means $ b $ of such diagrams. The generating functions for such numbers can be written in different ways. The simplest generating functions are proposed and investigated in \cite{Lee2014, MMN2019}. They satisfy both the 2BKP hierarchy and modified cut-and-join equations. We discuss these issues in the present paper.

\section{$Q$ Schur functions}

In this paper, we define the $Q$ Schur functions in a peculiar normalization,
which is conventional for fermionic representations,
but different from the natural one for the Cauchy identities and matrix models
used in \cite{MMq} and \cite{Alex}.

\subsection{Projective Schur functions}

To define the $Q$ Schur functions $Q_\alpha$, we  begin following \cite{Mac}  by defining an infinite skew
symmetric matrix $(Q_{ij})_{i, j \in \Nb}$, whose entries are symmetric functions of the infinite sequence
of indeterminates ${\bf x} =(x_1, x_2, \dots )$, via the following formula:
\be\label{Q_ij}
Q_{ij}({\bf x}) :=
\begin{cases}
q_i({\bf x}) q_j({\bf x}) + 2\sum_{k=1}^j (-1)^kq_{i+k}({\bf x}) q_{j-k}({\bf x}) \quad \text{if } (i,j) \neq (0,0), \\
0  \quad \text{if } (i,j) = (0,0) ,
\end{cases}
\ee
where the $q_i({\bf x})$'s are defined by the generating function:
\be
\prod_{i=1}^\infty {1+ z x_i \over 1 - z x_i}= \sum_{i=0}^\infty z^i q_i({\bf x})
\ee
For instance, $q_1(\xb)=2\sum_i x_i$. In particular,
\be
Q_{(j,0)}\left({\bf x}\right) = -Q_{(0,j)}\left({\bf x}\right)=q_j\left({\bf x}\right) \ \text{ for } j\ge 1
\ee

For a strict partition $\alpha$ of even cardinality $2n$ (including a possible zero part $\alpha_{2n}=0)$,
let $\Mb_{\alpha}({\bf x})$ denote the $2n\times 2n$ skew symmetric  matrix with entries
\be
\left(\Mb_{\alpha}({\bf x})\right)_{ij} := Q_{\alpha_i \alpha_j}({\bf x}), \quad 1\le i, j \le 2n.
\ee
The $Q$ Schur function is defined as its Pfaffian \cite{Mac}
\be\boxed{
Q_\alpha({\bf x}):= \Pf(\Mb_{\alpha}({\bf x}))}
\label{Q+_pfaff}
\ee
and, for completeness,
\be
Q_{\emptyset} :=1.
\ee

Equivalently,  these may be viewed as functions $q_j(\pb)$, $Q_{ij}(\pb)$
 of the  odd-indexed power sum symmetric functions $\pb=(p_1, p_3, \dots)$
\be
p_{2i-1} = p_{2i-1}({\bf x}) =  \sum_{a=1}^\infty x_a^{2i-1},  \quad a =1,2 , \dots.
\ee
Following \cite{Mac}, we use the agreement
\be
Q_\alpha\{p_k(\xb)\}:=Q_\alpha(\xb).
\ee
In particular, $Q_{[1]}\{p_k\}=2p_1=2\sum_i x_i$, $Q_{[2]}\{p_k\}=2p_1^2=2\left(\sum_i x_i\right)^2$.

There is the following relation:
\be\label{dR}
Q_\alpha\{\delta_{k,1}\}=
\frac{2^{|\alpha|}}{\prod_{i=1}^{\ell(\alpha)} \alpha_i!}\Delta^*(\alpha),\quad
 \Delta^*(\alpha):= \prod_{i<j\le \ell(\alpha)}
\frac{\alpha_i-\alpha_j}{\alpha_i+\alpha_j}
\ee

\br\label{BKP-higher-times}
In the literature on integrable systems, the variables often used and called times of the BKP hierarchy (see \cite{DJKM1},\cite{DJKM2},\cite{You},
\cite{O2003},\cite{NimmoOr},\cite{HvdLO}) are $\displaystyle{2p_m\over m}$ with $m$ odd.

In the present paper, we use the BKP hierarchy, however we re-write known BKP formulas in the power sum variables.
\er

\subsection{Neutral fermions and projective Schur functions}

In this section, we very briefly recall the known facts (details can be found
either in the original papers \cite{DJKM1},\cite{JM}, \cite{You}, or in \cite{O2003}, the results of which we will use).
Anyway, we need to fix the notation.

\subsubsection{From KP to BKP}

A natural way to construct the BKP hierarchy is to start with the KP hierarchy. A standard way to describe this latter is to realize the $\tau$-function of the hierarchy as a fermionic average
\be
\tau(\pb_f,\pb_f^*)=<0|\gamma(\pb_f)g\gamma^\dag(\pb_f^*)|0>
\ee
where $\pb_f:=(p_1,p_2,p_3,\ldots)$ and analogously $\pb_f^*$ are sets of KP time variables,
\be
{\gamma}({\pb_f}) :=  e^{\sum_{m>0} \frac1m {\cal J}_{ m} p_{m} },\quad
{\gamma}^\dag({\pb_f}) :=  e^{\sum_{m>0} \frac1m {\cal J}_{- m} p_{m} }\cr
{\cal J}_m:=\sum_{i\in\mathbb{Z}}\psi_i\psi^*_{i+m},\quad
g=\exp\left(\sum_{i,j}A_{ij}\psi_i\psi^*_j\right)
\ee
and $\psi_i$, $\psi_i^*$ are charged fermions,
\be\label{canonical-charges}
[\psi_i,\psi_j]_+=0,\quad [\psi_i,\psi_j^*]_+=\delta_{ij},\quad [\psi_i^*,\psi_j^*]_+=0
\ee
with the vacuum defined as
\be
\psi_i|0>=0=<0|\psi_i^*\ \ \ \ \forall i<0,\quad <0|\psi_i=0=\psi^*_i|0>\ \ \ \ \forall i\ge 0
\ee
The pairwise expectation values are:
\bea\label{psi-pairing}
    \langle 0| \psi^\dag_j \psi_k |0\rangle&\& =
    \begin{cases}
      \delta_{j,k}& \text{if}\ k>0,\\
      0& \text{if}\ k<0,
    \end{cases}
\eea

Now one can notice that an embedding into the KP hierarchy of the $\tau$-function that depends on only odd time variables can be naturally achieved by introducing the two sets of neutral fermions $\{\phi_i,\,i\in\mathbb{Z}\}$
\be\label{phi-psi}
\phi_j={\psi_j+(-1)^j\psi_{-j}^*\over\sqrt{2}},\quad \hat\phi_j=i{\psi_j-(-1)^j\psi_{-j}^*\over\sqrt{2}}
\ee
with the canonical anticommutation relations:
\be
\label{neutral-canonical}
 [\phi_j,\phi_k]_+ =(-1)^j \delta_{j+k,0},\quad  [\hat\phi_j,\hat\phi_k]_+ =(-1)^j \delta_{j+k,0},\quad  [\phi_j,\hat\phi_k]_+ =0
\ee
In particular,
$
(\phi_0)^2=\tfrac{1}{2}
$.
Acting on the left and right vacua $|0\rangle$, $\langle 0|$, one obtains
\be
\phi_{-j} |0\rangle  =0 =\langle 0| \phi_{j}, \quad  \forall  j > 0  .
\label{phi_vac_r}
\ee
and similarly for the second set of fermions. The pairwise expectation values are:
\bea\label{b-pairing}
    \langle 0| \phi_j \phi_k |0\rangle&\& =
    \begin{cases}
      (-1)^k\delta_{j,-k}& \text{if}\ k>0,\\
      \tfrac12\delta_{j,0}& \text{if}\ k=0,\\
      0& \text{if}\ k<0,
    \end{cases}
\eea

For (Euclidean) Fermi fields $ \phi(\zeta): = \sum_{n \in \mathbb {Z}} \zeta^n \phi_n $,
where $ \zeta \in \mathbb {CP}^1 $,
the variables $ T = - \log | \zeta | $ and $ \varphi = \arg \zeta $ are treated as time and space variables respectively.
The multipoint correlation functions uses the chronological ordering convention that states that the field assigned
to a larger time variable is put to the right of the field assigned to a smaller one, and the permutation sign should be
taken into account. For example, if $ | \zeta_1 |> | \zeta_2 | $, one writes
$\langle 0|\phi(\zeta_2)\phi(\zeta_1)0\rangle$ as $
- \langle 0|\phi(\zeta_1)\phi(\zeta_2)0\rangle$, which by (\ref{b-pairing})  is equal to
$ -\frac 12+ \zeta_2\zeta_1^{-1}-\cdots  =-\frac 12 \frac{1-\zeta_2\zeta_1^{-1}}{1+\zeta_2\zeta_1^{-1}}$.
For a given strict partition $\alpha=(\alpha_1,\dots,\alpha_{2r})$ where $\alpha_{2r}\ge 0$, we
use the notation $\Delta^*(\zeta_\alpha)$ defined by the relation
\be\label{multicorr}
\langle 0|\phi(\zeta_{\alpha_1})\cdots \phi(\zeta_{\alpha_{2r}})  |0\rangle =
2^{-r}\prod_{i<j\le 2r}\frac{\zeta_{\alpha_i}-\zeta_{\alpha_j}}{\zeta_{\alpha_i}+\zeta_{\alpha_j}}
=:2^{-r}
\Delta^*(\zeta_\alpha)
\ee
The first equality is obtained by the Wick theorem and by a simple analytical reasoning about poles and zeroes.

An important fact is that the factor $\gamma(\pb_f)$ becomes a product of two factors at all even times vanishing:
\be
\gamma(\pb_f)\Big|_{p_{2k}=0}= e^{\sum_{m\in\mathbb{Z}^+_{odd}} \left(\frac2m  p_{m}J_{ m}+\frac2m  p_{m}\hat J_{ m}\right) }
\ee
where
\be
J_m = \frac 12 \sum_{i\in\mathbb{Z}}(-1)^i:\phi_{-i-m}\phi_{i}:
\ee
and similarly for $\gamma^\dag(\pb^*)$. Here $:X:$ denotes the normal ordering (which is $:X:=X-\langle 0|X|0\rangle$ for $X$ quadratic in fermions).

Now we can consider only ``half" of this system leaving only one set of the neutral fermions.
In this system, there are  two mutually commuting Abelian groups of the BKP flows
\be
{\gamma}({\pb}) :=  e^{\sum_{m\in\mathbb{Z}^+_{odd}} \frac2m J_{ m} p_{m} },\quad
{\gamma}^\dag({\pb}) :=  e^{\sum_{m\in\mathbb{Z}^+_{odd}} \frac2m J_{- m} p_{m} }
\ee
One has
\be\label{HeisenbergB}
J_nJ_m-J_mJ_n = \frac n2\delta_{n+m,0}
\ee
and
\be\label{J-vac}
J_m|0\rangle = 0 =\langle 0|J_{-m},\quad \forall m >0
\ee
which results in
\be\label{gamma-vac}
\gamma(\pb)|0\rangle =|0\rangle ,\quad
\langle 0|\gamma^\dag(\pb)=\langle 0|
\ee
and in
\be
\gamma(\pb)\gamma^\dag(\pb^*)=e^{\sum_{m\in\mathbb{Z}^+_{odd}}\tfrac 2m p_mp^*_m   }
\gamma^\dag(\pb^*)\gamma(\pb)
\ee

\subsubsection{$Q$ Schur function as fermionic average}

One can construct the $Q$ Schur functions as fermionic averages much similar to how the ordinary Schur functions are realized as fermionic averages of charged fermions \cite{JM}.

Any nonzero partition  with distinct parts (also known as strict partition), say $\alpha$, can be written
as $\alpha=(\alpha_1,\dots,\alpha_r)$, where $r$ is an even number, and $\alpha_1>\cdots >\alpha_r\ge 0$.
We call $r$ the completed length of $\alpha$ and denote ${\bar\ell}(\alpha)$.
As usual, the length of a partition is the number of non-vanishing parts of $\alpha$, and it is denoted
$\ell(\alpha)$; thus $\ell(\alpha)$
is an odd number if and only if $\alpha_r=0$, while ${\bar\ell(\alpha)}$ is always even. The length of the zero partition is 0. Following \cite{Mac}, we denote by $\DP$
the set of all partitions with distinct parts (or the same: the set of all strict partitions).

Let us introduce the notation
\be
\Phi_\alpha=2^n\phi_{\alpha_1}\cdots \phi_{\alpha_{2n}},\quad
\Phi^\dag_\alpha =(-1)^{\sum_{i=1}^{2n}\alpha_i}2^{n}\phi_{\alpha_{-2n}}\cdots \phi_{-\alpha_1}
\nn\\
J^\dag_\Delta =J_{\Delta_1}\cdots J_{\Delta_m},\quad
J_\Delta =J_{-\Delta_m}\cdots J_{-\Delta_1}
\ee
where $\Delta=(\Delta_1,\dots,\Delta_m)$ is a partition with odd parts (we denote $\OP$ the total set of such partitions).
We have
\be
\langle 0|\Phi^\dag_{\beta}\Phi_\alpha |0\rangle = 2^{2n} \delta_{\alpha,\beta}=
<Q_\alpha,Q_\beta>
\nn\\
\langle 0|J^\dag_{\tilde{\Delta}}J_\Delta |0\rangle = 2^{-m}z_{\Delta}\delta_{\tilde{\Delta},\Delta}
=<\pb_{\tilde\Delta},\pb_\Delta>
\ee
where $<,>$ denotes the scalar product in the space of symmetric functions, see (8.12) in \cite{Mac}. Here  $z_\Delta$ is the standard symmetric factor of the Young diagram (order of the automorphism), and $\pb_\alpha:=\prod_i^{\ell(\alpha)}p_{\alpha_i}$.

The key relation we need was found in \cite{You} and, in our notations, is
\be\boxed{
 Q_\alpha\{p_k\}  =
 \langle 0| \gamma(\pb)\Phi_{\alpha} |0\rangle =
 \langle 0| \Phi^\dag_{\alpha} \gamma^\dag(\pb) |0\rangle}
 \label{Q+fermi}
\ee
which results in (\ref{Q+_pfaff}) according to the Pfaffian form of the Wick theorem with the choice $p_m = \sum_{i} x_i^m$.

As a result, one has
\be\label{coherent}
\gamma^\dag(\pb)|0\rangle =\sum_{\alpha\in\DP} 2^{-2n} \Phi_{\alpha} |0\rangle
Q_\alpha\{p_k\}
=\sum_{m\ge 0} \sum_{\Delta\in\OP\atop\ell(\Delta)=m} 2^{m} J_{\Delta}|0\rangle \frac{\pb_\Delta}{z_\Delta}
\nn\\
\langle 0|\gamma(\pb)=\sum_{\alpha\in\DP} 2^{-2n}
Q_\alpha\{p_k\}\,\langle 0|\Phi^\dag_{\alpha}
=   \sum_{k\ge 0} \sum_{\Delta\in\OP\atop\ell(\Delta)=k}    2^{m}  \frac{\pb_\Delta}{z_\Delta}
\langle 0|J^\dag_{\Delta}
\ee
thus, one gets
\be
e^{\sum_{m\in\mathbb{Z}^+_{odd}}\tfrac 2m p_mp^*_m   }=
\langle 0|\gamma(\pb)\gamma^\dag(\pb)|0\rangle
=\sum_{\alpha\in DP}
2^{-\ell(\alpha)} Q_\alpha\{p_k\}Q_\alpha\{p_k^*\}=
\sum_{\Delta\in\OP} 2^{\ell(\Delta)}\frac{\pb_\Delta \pb^*_\Delta}{z_\Delta}
\ee
where the first equality follows from (\ref{J-vac} and (\ref{gamma-vac})), see
also (8.13) in \cite{Mac}.

Let us write down the unity operator which acts in the Fock space spanned by $\Phi_\alpha |0\rangle$:
\be\label{unity}
{\bf 1} =\sum_{\alpha\in\DP} \Phi_\alpha |0\rangle 2^{-\ell(\alpha)} \langle 0|\Phi^\dag_\alpha
= \sum_{\Delta\in\OP} J_\Delta |0\rangle \frac{2^{\ell(\Delta)}}{z_\Delta} \langle 0|J^\dag_\Delta
\ee
The sets $\{\Phi_\alpha |0\rangle,\,\alpha\in\DP\}$ and $\{J_\Delta |0\rangle,\,\Delta\in\OP\}$
form two different bases in the fermionic Fock space.

\subsubsection{Sergeev characters}

Now note that the quantity
\be\boxed{
   \chi_\alpha(\Delta) :=2^{-\ell(\alpha)} \langle 0|J^\dag_{\Delta}\Phi_{\alpha} |0\rangle
=2^{-\ell(\alpha)} \langle 0|\Phi^\dag_{\alpha}J_{\Delta} |0\rangle}
\ee
is nothing but the character of the Sergeev group \cite{Serg}. To see this, we notice that
\be
J_\Delta |0\rangle =\sum_\alpha \chi_\alpha(\Delta)\Phi_\alpha |0\rangle ,\quad
\langle 0|J^\dag_\Delta = \sum_\alpha \chi_\alpha(\Delta) \langle 0|\Phi^\dag_\alpha
\nn\\
\Phi_\alpha |0\rangle  =\sum_\Delta \frac{2^{\ell(\alpha)+\ell(\Delta)}}{z_\Delta}
\chi_\alpha(\Delta)J_\Delta |0\rangle, \quad
\langle 0|\Phi^\dag = \sum_\Delta \frac{2^{\ell(\alpha)+\ell(\Delta)}}{z_\Delta}
\chi_\alpha(\Delta) \langle 0| J^\dag_\Delta
\ee
Thus, one has
\be
\pb_\Delta =\sum_\alpha \chi_\alpha(\Delta)Q_\alpha\{p_k\},\quad
Q_\alpha\{p_k\}=\sum_\Delta \frac{2^{\ell(\alpha)+\ell(\Delta)}}{z_\Delta}
\chi_\alpha(\Delta)\pb_\Delta
\ee
This is a counterpart of the Fr\"obenius formula for the $Q$ Schur functions (see \cite[Sec.I.7]{Mac}), and, hence, $\chi_\alpha(\Delta)$ are, indeed, characters of the Sergeev group, \cite{G,Lee2014}.

Introduce
\be
\f_\alpha(\Delta) :=
2^{-\ell(\Delta)}\langle 0|J^\dag_{\Delta}\Phi_{\alpha} |0\rangle \frac{1}{z_\Delta}\frac{1}{Q_\alpha\{\delta_{k,1}\}}
=\frac{2^{\ell(\alpha)+\ell(\Delta)}}{z_\Delta Q_\alpha\{\delta_{k,1}\} }
\chi_\alpha(\Delta)
\ee
In \cite{MMN2019}, we used the notation $\Phi_\alpha(\Delta)$ for this quantity, it is a basic building block for a spin counterpart of the Fr\"obenius formula for the Hurwitz numbers, see formula (\ref{Hur}).

With this quantity,
\be
Q_\alpha\{p_k\}=Q_\alpha\{\delta_{k,1}\}\sum_{\Delta\in\OP} \f_\alpha(\Delta) \pb_\Delta
\ee

\section{BKP $\tau$-functions}

Any vacuum expectation value of the form
\be
\tau(\pb,\pb^*)=\langle 0|\gamma(\pb) g \gamma^\dag(\pb^*) |0\rangle
\ee
where $\pb=(p_1,p_3,\dots)$ and $\pb^*=(p_1^*,p_3^*,\dots)$ are two independent sets of parameters, and
\be
g=e^{\sum_{i,k\in\mathbb{Z}} A_{ik}:\phi_i\phi_k:}
\ee
is a BKP $\tau$-function with respect to the set of times $ \pb  $, and, at the same time, it is a BKP
$\tau$-function with respect to the set of times $ \pb^* $, hence, it can be called a two-sided BKP, we call it either
BKP $\tau$-function or 2BKP $\tau$-function.
Here  $\{A_{ik}\}_{i,k \in \Zb}$ are the elements of a doubly infinite skew symmetric matrix $A$. The choice of
matrix $A$ or, what is the same, the choice of $g$ defines a common solution to all equations of the
2BKP integrable hierarchy.

\subsection{Hypergeometric BKP vs KP $\tau$-functions}

The term hypergeometric $\tau$-function was introduced in \cite{OS} where it was emphasized that the series
\be
\sum_{R} {\prod_{l=1}^p(a_l)_{_{R}}
\over\prod_{m=1}^s(b_m)_{_{R}}}
\cdot{\rm Schur}_R({\bf x})\cdot{\rm Schur}_R({\bf y}),\ \ \ \ \ \ \ \ \
(z)_R:=\prod_{(i,j)\in R} (z+j-i)
\ee
is a generalized hypergeometric series of two sets of variables ${\bf x}$ and ${\bf y}$.
This series can be presented in the form
\be\label{Cas}
\sum_R \hbox{Schur}_R\{p_k\}\hbox{Schur}_R\{p_k^*\}e^{\sum_k {1\over k}t_kC_k(R)},\ \ \ \ \ \ \ \
C_k(R):=\sum_i (R_i-i+1/2)^k-(1/2-i)^k
\ee
(see, e.g., \cite{AMMN2}), where $C_k(R)$ are the eigenvalues of the peculiarly chosen Casimir operators of $GL(\infty)$ in representation $R$ of $SL(N)$ at large enough $N> \ell(R)$. This is a KP $\tau$-function with $g$ of the form \cite{GKM2,OS,Ok}
\be
g=e^{\sum_{n\in\mathbb{Z}}:\ \psi_n\psi^*_n\ :\sum_k \frac 1k n^kt_k}
\ee
Quite similarly, the special choice of $g$,
\be\label{g}
g^\pm=g^\pm(\tb)=
e^{ \sum_{n>0}(-1)^n \phi_n\phi_{-n} \left(\sum_{k\in\mathbb{Z}^+_{odd}} \frac 1k n^k t_k -\delta^\pm\right)},
{   \quad
\delta^\pm = {i\pi(\pm 1-1)\over 2},}
\ee
where $\tb=(t_1,t_3,\dots)$ is a set of parameters, describes the family of $\tau$-functions called
hypergeometric BKP $\tau$-functions
\cite{O2003},\cite{NimmoOr}. We will denote the related $\tau$-function through
$\tau^\pm(\pb,\pb^*|\tb)$.

One can verify that
\be
 g^\pm  \phi_i \left( g^\pm \right)^{-1}
=\pm e^{ \sum_{k\in\mathbb{Z}^+_{odd}} \frac 1k i^k t_k }\phi_i,\quad i\neq 0
\ee
Let us note that $g^\pm | 0\rangle = | 0\rangle $.
From (\ref{neutral-canonical}), (\ref{coherent}), it follows that
\be\label{VEV1}
\tau^\pm(\pb,\pb^*|\tb):=\langle 0|\gamma(\pb) g^\pm(\tb) \gamma^\dag(\pb^*) |0\rangle=
\ee
\be
 = \sum_{\alpha\in \DP\atop \ell(\alpha)\,{\rm even}} e^{\sum_{k\in\mathbb{Z}^+_{odd}} \frac 1k t_k \omega_k(\alpha)   }
2^{-\ell(\alpha)} Q_\alpha(\pb)Q_\alpha(\pb^*)
 \pm
\sum_{\alpha\in \DP\atop \ell(\alpha)\,{\rm odd}} e^{\sum_{k\in\mathbb{Z}^+_{odd}} \frac 1k t_k \omega_k(\alpha) }
2^{-\ell(\alpha)} Q_\alpha(\pb)Q_\alpha(\pb^*):\nn
\ee
i.e.
\be\label{pm-hyp-tau}\boxed{
\tau^\pm(\pb,\pb^*|\tb)=\hbox{R}_\pm \cdot\sum_{\alpha\in \DP} e^{\sum_{k\in\mathbb{Z}^+_{odd}}
\frac 1k t_k \omega_k(\alpha) }
2^{-\ell(\alpha)} Q_\alpha(\pb)Q_\alpha(\pb^*)}
\ee
where $\hbox{R}_\pm$ is the projection operator, and
\be
\omega_k(\alpha) = \sum_{i=1}^{\ell(\alpha)}\alpha_i^k
\ee
are spin counterparts of the completed cycles in the case of ordinary Hurwitz numbers. Notice the difference between these quantities and (\ref{Cas}): in the spin case, the main quantities involve the integers $\alpha_i$, while the same quantities in the non-spin case, $\alpha_i-i$. This is related to the fact that, in the former case, all the formulas involve the strict partitions, and, in the latter one, the formulas involve all partitions.

\br
 For every hypergeometric BKP $\tau$-function (\ref{pm-hyp-tau}), its square is a hypergeometric KP $\tau$-function,
with all even times put to zero,  details see in \cite[sec.3,Prop.2]{O2003}.
\er

The issue of the hypergeometric $\tau$-functions can be explained in a different way: the KP $\tau$-function can be generally expanded into the Schur functions,
\be\label{Sex}
\tau(\pb)=\sum_\alpha c_\alpha{\rm Schur}_\alpha\{p_k\}
\ee
This linear combination solves the KP hierarchy iff the coefficients satisfy the Pl\"ucker relations for the infinite-dimensional Grassmannian \cite{JM}
\be\label{Plucker}
c_{(\vec x|\vec y)[x_i,x_j;y_i,y_j]}\cdot c_{(\vec x|\vec y)}-
c_{(\vec x|\vec y)[x_i;y_i]}\cdot c_{(\vec x|\vec y)[x_j;y_j]}+
c_{(\vec x|\vec y)[x_i;y_j]}\cdot c_{(\vec x|\vec y)[x_j;y_i]}=0
\ee
where we used the Fr\"obenius (hook) parametrization of the Young diagrams $\alpha=(x_1,\ldots,x_h|y_1,\ldots,y_h)=(\vec x|\vec y)$, where $x_i,y_i$ are the hook arms and legs correspondingly \cite{Mac}, and denoted through $[\{x_i\};\{y_j\}]$ removing a subset $\{x_i\};\{y_j\}$ from the set of hook legs and arms.

Now we note that if $c_\alpha$'s solve these relation, so do $c_\alpha\prod_{i,j\in\alpha}f(i-j)$ with an arbitrary function $f(x)$. Since a particular solution to the
Pl\"ucker relations is given by the Schur functions of arbitrary set of times (parameters), and since the exponential in (\ref{Cas}) is of the type $\prod_{i,j\in\alpha}f(i-j)$ \cite{AMMN2}, we finally come to
the hypergeometric $\tau$-function (\ref{Cas}).

Similarly, a linear combination of the $Q$ Schur functions,
\be\label{Qex}
\tau(\pb)=\sum_{\alpha\in\DP} c^{BKP}_\alpha Q_\alpha\{p_k\}
\ee
solves the BKP hierarchy iff the coefficients satisfy the Pl\"cker relations for the isotropic Grassmannian \cite{JM}
\be
c^{BKP}_{[\alpha_1,\ldots,\alpha_k]}c^{BKP}_{[\alpha_1,\ldots,\alpha_k,\beta_1,\beta_2,\beta_3,\beta_4]}
-c^{BKP}_{[\alpha_1,\ldots,\alpha_k,\beta_1,\beta_2]}c^{BKP}_{[\alpha_1,\ldots,\alpha_k,\beta_3,\beta_4]}+\nn\\
+c^{BKP}_{[\alpha_1,\ldots,\alpha_k,\beta_1,\beta_3]}c^{BKP}_{[\alpha_1,\ldots,\alpha_k,\beta_2,\beta_4]}
-c^{BKP}_{[\alpha_1,\ldots,\alpha_k,\beta_1,\beta_4]}c^{BKP}_{[\alpha_1,\ldots,\alpha_k,\beta_2,\beta_3]}=0
\ee
Now we again note that if $c^{BKP}_\alpha$'s solve these relation, so do $(\pm 1)^{\ell(\alpha)}e^{\sum_{m\in\mathbb{Z}^+_{odd}} t_m \omega_m(\alpha) } c^{BKP}_\alpha$. Since a particular solution to the Pl\"ucker relations is given by the $Q$ Schur functions of arbitrary set of times (parameters), we finally come to the hypergeometric $\tau$-function (\ref{pm-hyp-tau}).

\subsection{Bosonization}

Now we need to bosonize
the operators that act on the Fock vectors\footnote{ Note that, for the first time, such a bosonization relation was obtained in an unpublished preprint preceding the article \cite{PogrebkovSushko}, the article was not accepted for publication because the result was rather unusual.  }.
The BKP hierarchy and related objects that we will review in this section were introduces in a series of papers by
Kyoto school \cite{DJKM1},\cite{DJKM2},\cite{JM}. However, we shall use here the approach due to \cite{Leur1994},\cite{KvdLbispec}.

Let $z\in S^1$
\be\label{vertex-B}
V(z,\hat{\pb})=\tfrac{1}{\sqrt {2}}D_\eta
e^{\sum_{m\in\mathbb{Z}^+_{odd}} \tfrac2m z^m p_m}e^{-\sum_{m\in\mathbb{Z}^+_{odd}}  z^{-m} \tfrac {\partial}{\partial p_m}}
\ee
be the vertex operator as it was introduced in \cite{KvdLbispec}. Here
$D_\eta=\eta+\frac{\partial}{\partial\eta}$, where $\eta$ is an auxiliary odd Grassmannian variable:
$\eta^2=0$, $D_\eta^2=1$. The symbol ${\hat{\pb}}$ denotes the set of two collection $p_1,p_3,p_5,\dots$
and $\tfrac {\partial}{\partial p_1},\tfrac {\partial}{\partial p_3},\dots$.

Introducing
\be
2\theta(z,\pb) := \sum_{m\in\mathbb{Z}^+_{odd}} \tfrac 2m  z^m p_m
- \sum_{m\in\mathbb{Z}^+_{odd}}   z^{-m} \tfrac {\partial}{\partial p_m} =\sum_{m\in\mathbb{Z}_{odd}} \frac 2m J^b_m z^m
\ee
where
\be
J^b_m(\hat\pb) =\begin{cases}
               p_m \ \text{ if } \ m>0\ \ \text{odd} \cr
             -\frac m2 \tfrac {\partial}{\partial p_{-m}}        \ \text{ if } \ m<0\ \ \text{odd} \cr
           0\ \text{ if } \ m \ \text{even},
          \end{cases}
\ee
one can rewrite (\ref{vertex-B}) as
\be
V(z,\pb)=\tfrac {D_\eta}{\sqrt{2}}\vdots e^{2\theta(z,\pb)} \vdots
\ee
where $\vdots\vdots$ denotes the so-called bosonic normal ordering which means that all derivatives are moved
to the right.

Using
\be
[J^b_m,J^b_n]=-\frac{m}{2}\delta_{m+n,0},
\ee
one can verify that
\be
V(z_1,\pb)V(z_2,\pb)=\frac{1-z_2z_1^{-1}}{1+z_2z_1^{-1}}\vdots V(z_1,\pb)V(z_2,\pb) \vdots
\ee
which results in the commutation relations
\be
V(z_1,\pb)V(z_2,\pb)+V(z_2,\pb)V(z_1,\pb)=\sum_{n\in\mathbb{Z}}\left(-\frac{z_1}{z_2}  \right)^n  =
\delta(\varphi_1-\varphi_2-\pi)
\ee
where the right hand side is the Dirac delta function, $z_i=e^{\varphi_i},\,i=1,2$.
For the Fourier modes $V(z,\pb)=\sum_{i\in\mathbb{Z}} V_i(\pb) z^i$, one obtains
\be
V_n(\pb)V_m(\pb) +V_m(\pb)V_n(\pb) = (-1)^n \delta_{n+m}
\ee
Now one can note that up to the sign factor the fermionic and bosonic currents have the same commutation relations:
\be\label{currents-commutator}
J^b_n J^b_m - J^b_m J^b_n
= J_m J_n - J_n J_m = \frac m2 \delta_{n+m,0}
\ee
and that the operator
\be
\phi(z): = \phi_0 \gamma([z]) \gamma^\dag(-[z^{-1}])
\ee
where $[z]:=(z,z^3,z^5,\dots)$,
has the same commutation relations as the vertex operator (\ref{vertex-B}). This describes a correspondence between bosonic and fermionic operators.

\subsection{$W^B_\infty$ algebra \label{W}}

Following the standard procedure, one can expand the product of the two vertex operators in the generators of $W^B_\infty$ algebra:
\be
\frac 12 V(ze^{\tfrac y2},\pb)   V(-ze^{-\tfrac y2},\pb) - \frac 12 \frac{1+e^{-y}}{1-e^{-y}} =
\frac 14 \frac{e^y+1}{e^y-1}\left(\vdots e^{\theta(ze^{\tfrac y2})+\theta(-ze^{-\tfrac y2})} \vdots -1\right)=
\nn\\
=\frac 14 \frac{e^y+1}{e^y-1}
\sum_{k>0}\frac{1}{k!}\vdots \left(
\sum_{m\in\mathbb{Z}_{odd}}\theta_m z^m \left(e^{\tfrac {my}{2}} -e^{-\tfrac {my}{2}}\right)
\right)^k \vdots
=:\sum_{m\in\mathbb{Z},n\ge 0} \frac {1}{n!} y^{n} z^m \Omega_{mn}(\pb)
\label{parted-vertices}
\ee
As follows from the left hand side of this formula, $\Omega_{mn}$ vanish when $n$ and $m$ have
the same parity.

In particular, the operators $\Omega_{m,0}=J^b_m$ with odd $m$ form the bosonic current algebra,
$\Omega_{m,1}$ with even $m$ form the Virasoro algebra, etc. Our main interest is the commutative
algebra of operators $\Omega_n:= \Omega_{0,n}$ with odd $n$. In particular,
$$
\Omega_1=\sum_{n>0} n p_n\partial_n
$$
\be
\Omega_3 = \frac 12 \sum_{n>0} n^3 p_n\partial_n +
\frac {1}{2}\sum_{n>0} n p_n\partial_n
+4\sum_{n_1,n_2,n_3\,{\rm odd}} p_{n_1}p_{n_2}p_{n_3}  (n_1+n_2+n_3)\partial_{n_1+n_2+n_3} +
\nn\\
+3\sum_{n_1+n_2=n_3+n_4\,{\rm odd}} p_{n_1}p_{n_2} n_3n_4\partial_{n_3}\partial_{n_4} +
\sum_{n_1,n_2,n_3\,{\rm odd}} p_{n_1+n_2+n_3}\partial_{n_1}\partial_{n_2}\partial_{n_3}
\ee

The fermionic counterpart of (\ref{parted-vertices}) is much simplier:
\be
 \frac 12 :\phi(ze^{\tfrac y2})\phi(-ze^{-\tfrac y2}): = \frac 12
\sum_{m,j\in\mathbb{Z}} z^{m} e^{\tfrac y2(m+2j)}(-1)^j:\phi_{m+j}\phi_{-j}: =
\sum_{m\in\mathbb{Z}\atop n\ge 0} \frac{1}{n!}  y^n z^m \Omega^\textsc{f}_{mn}
\ee
Again, as follows from the left hand side of this formula, $\Omega_{mn}=0$ when $n$ and $m$ have the same parity.
One gets
\be
\Omega^\textsc{f}_{mn}=\frac 12 \sum_{j\in\mathbb{Z}}  \left(\tfrac m2+j\right)^n(-1)^j:\phi_{m+j}\phi_{-j}:
\ee

In particular, the current algebra:
\be
J_m:=  \Omega^\textsc{f}_{-m,0}=
\begin{cases} \frac 12 \sum_{j\in\mathbb{Z}}(-1)^j \phi_{j-m}\phi_{-j} \ \text{ if } \ m\ \ \text{odd} \cr
                  0  \ \text{ if } \ m\ \ \text{even},
\end{cases}
\ee
the Virasoro algebra
\be
L^\textsc{f}_m:=\Omega^\textsc{f}_{-m,1}=
\begin{cases} \frac 12 \sum_{j\in\mathbb{Z}} \left(\tfrac m2+j\right)(-1)^j
(-1)^j \phi_{j-m}\phi_{-j} \ \text{ if } \ m\ \ \text{even} \cr
                  0  \ \text{ if } \ m\ \ \text{odd},
\end{cases}
\ee
and
\be
\Omega^\textsc{f}_{n}:=\Omega^\textsc{f}_{0,n}=
\frac 12 \sum_{j\in\mathbb{Z}}(-1)^j j^n:\phi_{j}\phi_{-j}:=
\sum_{j=1,3,\dots}(-1)^j j^n\phi_{j}\phi_{-j},\quad n\,{\rm odd}.
\ee
One can easily see that
$$
[\Omega^\textsc{f}_{n},\Omega^\textsc{f}_{m}]=0
$$
for each pair of $n,m$.

The boson-fermion correspondence gives rise to the following relation
$$
e^{t_{mn}\Omega_{m,n}( \pb + t_{m'n'})\Omega_{m',n'}(\pb)} \cdot \tau(\pb,\pb^*)=
\langle 0| \gamma(\pb) e^{ t_{mn} \Omega^\textsc{f}_{m,n}} g  e^{ t_{m'n'} \Omega^\textsc{f}_{m',n'}} \gamma^\dag(\pb^*)|0\rangle
=\tau(\pb,\pb^*|t_{mn},t_{m'n'})
$$
The flows with respect to the parameters $t_{mn}$ and $t_{m'n'}$ are called additional symmetries,
see \cite{Orlov1988} and \cite{Leur1994}.

\bl
For $r\in\mathbb{Z}_{\ge 0}$ introduce
\be\label{Omega=res}
 \quad d_{2k,n} := z^{-k}\cdot\left(z\frac{\partial}{\partial z}\right)^n\cdot z^{-k}
\ee
$$
\Omega^{\textsc{f}}_{2k,n}(r):=\res_z \left(d_{2k,n}^r\cdot\phi(z)\right) \phi(-z)\frac {dz}{z}
=\sum_{j} \phi_{2rk+j} \phi_{-j}(-1)^j (j+k)^n(j+3k)^n \ldots (j+(2r+1)k)^n
$$
Then:
\begin{itemize}
\item $\Omega_{2k,1}(1)=\Omega_{2k,1}$ and  $\Omega_{2k,1}(r)=0$ if $r$ is even.
\item
$
[\Omega_{2k,n}(r),\Omega_{2k,n}(r')]=0.
$
\end{itemize}
\el
We define
\be\label{gamma}
\gamma_{m,n}(\pb):=e^{\sum_{r>0,{\rm odd}} p_r \Omega^\textsc{f}_{m,n}(r) }
\ee
where $\pb=(p_1,p_3,\dots)$ is a set of parameters.

It can be shown that the vacuum expectation value
$$
\langle 0|\gamma_{m,n}(\pb) g \gamma_{m',n'}(\pb^*) |0\rangle
$$
where $m<0$ and $m'>0$ is a BKP $\tau$-function with respect to $\pb$ variables and a BKP $\tau$-function
with respect to $\pb^*$ variables.

Consider the Virasoro element
$$
\Omega_{2,1}=L_{-1} =\frac 12 \sum_{j\in\mathbb{Z}} \phi_{2+j}\phi_{-j}(-1)^j(1+j)
$$
and the set of related commuting operators
\be\label{L_(-1)-based-series}
\Omega_{2,1}(r)=:L_{-1}(r)=
\frac 12 \res_z \Bigg(\underbrace{ \left({\partial\over\partial z}\cdot{1\over z}\right)^r}_{d_{2,1}^r}\phi(z)  \Bigg)\phi(-z){dz\over z}=
\ee
\be
=\frac 12 \sum_{j\in\mathbb{Z}} \phi_{2r+j}\phi_{-j}(-1)^j(1+j)(3+j)
\cdots (2r+j-1),
\quad r=1,3,5,\dots
\ee
We have the following counterpart of (\ref{Q+fermi})
\begin{proposition}
\be\label{Q-via-Vir}
\langle 0|\Phi^\dag_\alpha \gamma_{2,1} (\pb) |0\rangle =\begin{cases}
Q_\alpha(\pb)\prod_i \alpha_i !! \quad{\rm each}\,\alpha_i\,{\rm is}\,{\rm even}\cr
0\qquad\qquad {\rm otherwise}
\end{cases}
\ee
\end{proposition}
The proof is achieved with the observation that $\phi_{-j}(\pb):=
\left(\gamma_{2,1}(\pb)\right)^{-1}\phi_{-j} \gamma_{2,1} (\pb)$
is a finite linear combination of the modes $\phi_{-i}$ where $1\le i\le j$ and where $i$ is odd in case
$j$ is an odd  positive
number. Then, it follows that the presence of the odd parts leads to zero vacuum expectation value.
We take into account that $\langle 0|\gamma_{2,1}=\langle 0|$  and apply the Wick theorem
for $\phi_{\alpha_i}(\pb)$ and use the definition (\ref{Q+_pfaff}) of the projective Schur functions
to get the right hand side of (\ref{Q-via-Vir}).

\bc
 Let $g=\exp \sum_{n>0} \phi_i\phi_{-i}\log F(i)$ where $F$ is a function on the lattice $\mathbb{Z}_{>0}$.
 We have
 \be\label{Virasoro-Kontsevich-BGW-Fermi}
\langle 0|\gamma(\pb) g \gamma_{2,1}(\pb^*) |0\rangle  =
\sum_{\alpha\in\DP} Q_{2\alpha}(\pb)Q_\alpha(\pb^*) \prod_{i=1}^{\ell(\alpha)} F(2\alpha_i)\alpha_i!!
 \ee
\ec
\bc
Let
$\gamma^{\rm B}_{2,1}(\pb^*;\hat\pb)$ be the bosonic version of $\gamma_{2,1}(\pb^*)$. Then
\be\label{Virasoro-Kontsevich-Bose}
\gamma^{\rm B}_{2,1}(\pb^*;\hat\pb) \cdot 1  =
\sum_{\alpha\in\DP} Q_{2\alpha}(\pb)Q_\alpha(\pb^*) \prod_{i=1}^{\ell(\alpha)} \alpha_i!!
 \ee

 Let us note that the right hand side of (\ref{Virasoro-Kontsevich-BGW-Fermi}) has the form of the
 right hand side of (\ref{Kon}) and (\ref{BGW}).

\ec

\br
The perturbation series for the partition functions of the well-known $N\times N$ two-matrix model (and therefore
also one-matrix model) are
obtained from the
similar construction \cite{OS}. Let us consider the set of operators commuting with the KP Virasoro generator
$L_{-1}$:
\be
L^{\rm KP}_{-1}(r):=\res_z \left( \frac{d^r\psi(z)}{dz^n} \right)\psi^\dag(z),\quad r=1,2,3,\dots.
\ee
Then
\be
e^{\sum_{r>0} \frac{1}{r}p^*_r L^{\rm Bos}_{-1}(r)}\cdot 1 =
\sum_{\lambda} (N)_\lambda s_\lambda(\pb)s_\lambda(\pb^*)
\ee
which is a counterpart of (\ref{gamma}) for $\gamma_{2,1}(\pb)$, (\ref{L_(-1)-based-series}), and (\ref{Virasoro-Kontsevich-Bose}).
\er

At the end of this section we note that the action of the bosonic version of $\gamma_{m,n}$ with even $n$
(and odd $m$) on 1
always gives a version of the BKP hypergeometric tau function.

\subsection{Cut-and-join equation}

A specific form of the exponential means that the BKP $\tau$-function (\ref{pm-hyp-tau}) is a generating function of the Hurwitz numbers corresponding to the completed cycles. In the case of the ordinary Hurwitz numbers, one could consider the generating function of the simplest double Hurwitz numbers with two branching profiles fixed and all other ramifications being just double ramification points, $H_d(0|\Delta,\Delta^*,\Gamma_2^b)$. This generating function was a (KP) $\tau$-function \cite{Ok}. On the contrary, in the spin case, even the simplest generating function of
the spin Hurwitz numbers $H^{\pm}(\Gamma_d,\Delta^1,\Delta^2)$ where
\be\label{gammad}
\Gamma_d = (3,1^{d-3})
\ee
is not a $\tau$-function.

Indeed, the Hurwitz numbers can be represented as \cite{G,Lee2014,MMN2019}
\be\label{Hur}
H^\pm(\Delta^1,\dots,\Delta^k)=\hbox{R}_\pm\cdot\sum_{\alpha\in\DP}
\left( Q_\alpha\{\delta_{k,1}\}\right)^2  \f_\alpha(\Delta^1)\cdots \f_\alpha(\Delta^k)
\ee
and (see \cite[Eq.(102) and derivation in sec.6]{MMN2019})
\be\label{f3}
\f_\alpha(\Gamma_d) =\frac13 \omega_3(\alpha) - \left(\omega_1(\alpha)\right)^2 +\frac 23 \omega_1(\alpha),\quad
|\alpha|=d\ge 3
\ee
One can see that this expression is not a linear combination of the completed cycles $\omega_k(\alpha)$. However, we can still derive an equation for $\tau^\pm(\pb,\pb^*|{\bf t})$, which is a counterpart of the celebrated cut-and-join equation \cite{GJW}.

To this end, with the help of (\ref{f3}), we re-write (\ref{pm-hyp-tau}) as follows
\be\label{tau+-}
\tau^\pm(\pb,\pb^*|{\bf t})
=
\hbox{R}_\pm\cdot\sum_{\alpha\in\DP}  2^{-\ell(\alpha)}e^{t_1d+t_3\left(\f_\alpha(\Gamma)) +d^2 -\tfrac 23 d\right)}
e^{\sum_{n>3,{\rm odd}} \tfrac 1n t_n \omega_n(\alpha)}
Q_\alpha\{p_k\}Q_\alpha\{p_k^*\}=
\nn\\
= c\sum_{d\ge 0}  e^{t_1(d-\frac13)+t_3\left(d - \frac13\right)^2}
\Phi^\pm_d(\pb,\pb^*|t_3,t_5,\dots)
\ee
where $c = e^{\frac13 t_1 - \frac 19 t_3}$ and where
\be
\Phi^\pm_d(\pb,\pb^*|t_3,t_5,\dots)=
\hbox{R}_\pm\cdot\sum_{\alpha\in\DP\atop |\alpha|=d}e^{t_3 \f_\alpha(\Gamma)}2^{-\ell(\alpha)}
e^{\sum_{n>3,{\rm odd}} \tfrac 1n t_n \omega_n(\alpha)}
Q_\alpha\{p_k\}Q_\alpha\{p_k^*\}
\ee
Let us put $t_i=0,\,i>3$.
Then, we get
\be
\Phi^\pm_d(\pb,\pb^*|t_3)=\sum_{\Delta^1,\Delta^2\atop |\Delta^1|=|\Delta^2|}
\sum_{b\ge 0} \frac{t_3^b}{b!}
H^\pm\left(\Gamma_d^b,\Delta^1,\Delta^2\right)\pb_{\Delta^1}\pb_{\Delta^2}^*
\ee
where
\be
H^\pm\left(\Gamma^b_d,\Delta^1,\Delta^2\right) =
\hbox{R}_\pm\cdot\sum_{\alpha\in\DP\atop |\alpha|=d} \left( Q_\alpha\{\delta_{k,1}\}\right)^2
\left(\f_\alpha(\Gamma)\right)^b \f_\alpha(\Delta^1) \f_\alpha(\Delta^2)
\ee

Since the BKP $\tau$-function is defined up to a constant factor $c$, we obtain
\bt Multiply the BKP $\tau$-function (\ref{pm-hyp-tau}) with the factor $\frac 1c=e^{-\tfrac 13 t_1 +\tfrac 19 t_3}$. Then, one gets the cut-and-join equation
in form
\be
\left(  \frac{\partial}{\partial t_3} -
\left(\frac{\partial}{\partial t_1} \right)^2  \right) \cdot \tau^\pm(\pb,\pb^*|\tb)
={\cal W}\cdot\tau^\pm(\pb,\pb^*|\tb)
\ee
or, which is the same
\be
 \frac{\partial\Phi^\pm_d(\pb,\pb^*|t_3)}{\partial t_3} =
 {\cal W}\cdot\Phi^\pm_d(\pb,\pb^*|t_3)
\ee
where the cut-and-join operator
\be
{\cal{W}}=\tfrac 13 \Omega_{3}(\pb)-\Omega_{1}(\pb)\Omega_{1}(\pb)+\tfrac 23 \Omega_{1}(\pb)
=-\left(\sum_{n>0} np_n\partial_n \right)^2 +\left(\frac 23+\frac 16\right) \sum_{n>0} np_n\partial_n+\nn\\
+\frac 16 \sum_{n>0} n^3 p_n\partial_n
+\frac 43 \sum_{n_1,n_2,n_3\,{\rm odd}} p_{n_1}p_{n_2}p_{n_3}  (n_1+n_2+n_3)\partial_{n_1+n_2+n_3} +
\nn\\
+\sum_{n_1+n_2=n_3+n_4\,{\rm odd}} p_{n_1}p_{n_2} n_3n_4\partial_{n_3}\partial_{n_4} +
\frac 13\sum_{n_1,n_2,n_3\,{\rm odd}} p_{n_1+n_2+n_3}\partial_{n_1}\partial_{n_2}\partial_{n_3}
\ee

\et
Thus, the cut-and-join equation has a form of the quantum heat equation with the potential ${\cal{W}}$ .

By analogy with the case of the ordinary Hurwitz numbers analyzed by Boris Dubrovin in \cite{Dubr}, we have
\br
Each $\Omega_n(\pb)$ can be treated as a Hamiltonian of the quantum dispersionless modified KdV equation
on the circle
with the eigenstates given by $Q_\alpha( \pb)$ and the eigenvalues given by $\omega_n(\alpha)$.
If one introduces the Plank constant $\hbar$ and puts $v=\hbar^{-\frac12 }\sum_n e^{in\varphi }J_n^b$, then
the first nontrivial Hamiltonian $\hbar^2\Omega_3$ is
\be
{\cal{H}}_3 =
\int_0^{2\pi} d\varphi \vdots\left(v^4 + \hbar \left( v_\varphi \right)^2 \right)\vdots d\varphi
\ee

\er

\section{KdV soliton solution as the BKP solutions}

\subsection{Solitons of the KdV and BKP hierarchies which generate Hurwitz numbers\label{KdV}}

So far we discussed integrable properties with respect to time variables $\pb$ that are the variables of the $Q$
Schur functions. In this subsection, we discuss a relation to the soliton solution in $t_n$. That is, we want to
show that if we fix both sets $\pb$, $\pb^*$ so that they are equal to $\pb_1:=(1,0,0,\ldots)$, then
$\tau^\pm(\pb_1,\pb_1|\tb)$ turns into a recognizable soliton $\tau$-function of the KdV equation, where the role
of times is played by the set $\tb=(t_1,t_3,\dots)$.

\bt
Consider the $\tau$-function (\ref{tau+-}) where we restrict the times to be $\pb=\pb^*=\pb_1$ and
introduce
\be
u^\pm(\tb)=2\frac{\partial^2}{\partial t_1^2}\log\tau^\pm(\pb_1,\pb_1|\tb)
\ee
Then each $u^\pm$ is the  ($\infty$)-soliton solution of the KdV hierarchy with respect to the times
$\tb=(t_1,t_3,\dots)$. In particular,
\be
12u^\pm_{t_3}= u^\pm_{t_1t_1t_1}+ 6u^\pm u^\pm_{t_1}
\ee
If one puts all $t_i=0$ for $i>3$ and denote $(t_1,t_3)=(x,t)$, so that $(x,t)$ are the space-time coordinates in the
standard KdV theory \cite{Novikov-ed}, then
\be\label{Triple-Hur-KdV}
\tau^\pm(\pb_1,\pb_1|x+\log q ,t,0,0,\dots)=
 c\sum_{d \ge 3} q^d e^{x d+t\left(d^2 -\tfrac 23 d\right)}
 \sum_{b\ge 0} \frac{t^b}{b!}
H^\pm\left(\Gamma^b_d\right)
\ee

\et

Proof. This solution is well-known: it is a multi-soliton $\tau$-function. Indeed,
as it follows from (\ref{pm-hyp-tau}), (\ref{dR}) we have
\be\label{KdV-sol}
\tau^\pm(\pb_1,\pb_1|\tb)=1+\sum_{k=1}^\infty
\sum_{1\le \alpha_1<\cdots<\alpha_k} 2^{2|\alpha|-k}
\prod_{i<j\le k}\left(\frac{\alpha_i-\alpha_j}{\alpha_i+\alpha_j}\right)^2
\prod_{i=1}^k
\frac{e^{\eta(\alpha_i,\tb) +\delta^\pm}}{ (\alpha_i!)^2}
\ee
where $\delta^\pm$ is defined in (\ref{g}) and where
\be\label{eta}
 \eta(x,\tb):=\sum_{m=1,3,5,\dots} \frac 1m t_m x^m
\ee
which a special case of the general soliton solution of KdV hierarchy given by
\be
\tau_{\rm KdV}^{{\rm sol}\pm}(\tb)=
1+\sum_{k>0}\sum_{\alpha\in\DP\atop \ell(\alpha)=k} \left( \Delta^*(\zeta_\alpha) \right)^2
\prod_{i=1}^k e^{\eta(\alpha_i,\tb)+a_i-\delta^\pm}
\ee
\be\label{sum-over-k}
=
1+\sum_{i}e^{\eta^\pm_i}+
\sum_{i<j}\frac{(\zeta_i-\zeta_j)^2}{(\zeta_i+\zeta_j)^2}e^{\eta^\pm_i+\eta^\pm_j}+
\sum_{i<j\le k}\frac{(\zeta_i-\zeta_j)^2(\zeta_i-\zeta_k)^2(\zeta_j-\zeta_k)^2}{(\zeta_i+\zeta_j)^2
(\zeta_i+\zeta_k)^2(\zeta_j+\zeta_k)^2}e^{\eta^\pm_i+\eta^\pm_j+\eta^\pm_k}
+\cdots
\ee
where $\eta^\pm_j:=\eta(\zeta_j,\tb)+a_j-\delta^\pm$ and
\be\label{Delta(zeta)}
\Delta^*(\zeta_\alpha):=\prod_{i<j\le k}\frac{\zeta_{\alpha_i}-\zeta_{\alpha_j}}{\zeta_{\alpha_i}+\zeta_{\alpha_j}}
\ee
We recall that the parameters $\zeta_j$ play the role of the soliton momenta, the parameters $a_j$
define initial positions of solitons, these are the solitons
$u^+  \sim \frac 12\zeta_j^2\ch^{-2}(\frac 12 \zeta_j x + \frac 16\zeta_j^3 t + \frac 12 a_j)$ for the $\delta^+$ case,
and the solitons $u^-  \sim \frac 12\zeta_i^2 \sinh^{-2}(\frac 12 \zeta_i x + \frac 16\zeta_i^3 t+ \frac 12 a_i)$
for the $\delta^-$ case. Factors (\ref{Delta(zeta)}) describe interactions of solitons.

Choosing $\zeta_j= j,\,j=1,2,3,\dots$
and $a_j=-\log 2^{1-2j}(j!)^2$, and taking into account (\ref{dR}), (\ref{pm-hyp-tau})
one obtains (\ref{KdV-sol}).
Formula (\ref{Triple-Hur-KdV}) translates the sum over the number of solitons in the formula
(\ref{sum-over-k}) into a sum over momenta $d=\sum_k \alpha_{k}$, $d$ playing the role of the number of sheets
in the covering problem related to $H(\Gamma_d^b)$.

\br\label{How-to-get-H(Delta)}
To generate the Hurwitz numbers for the $d$-sheeted coverings, it is enough to consider the $N$-soliton
KdV $\tau$-function with
the parameters $\zeta_i$ filling the string $0,1,\dots,d$ where $d$ does not exceed the number of solitons $N$.
Such a $\tau$-function is holomorphic
 in the $(x,t)$-variables, and one can write
\be
H^\pm\left(\Gamma^b_d\right)= (2\pi)^{-1}\res_{t=0} t^{-b-1}
e^{-t\left(d^2 +\tfrac 23 d\right)}\int_{0}^{2\pi }dx e^{-ixd}  \tau_N^{{\rm KdV},\pm}(ix,t)
\ee
\er

A similar solitonic $\tau$-function can be obtained at $\pb^*=\delta_{k,1}:=\pb_1$:
$\tau(\pb,\pb_1|\tb)$ is the $\tau$-function of the two component BKP hierarchy \cite{KvdLbispec}
(or of the Veselov-Novikov hierarchy)
with respect to the times $\pb=(p_1,p_3,\dots)$ and $\tb=(t_1,t_3,\dots)$.
 \br
Two component $\tau$-functions are constructed with the two-component fermions.
We recall that any two-component BKP (2-BKP)
$\tau$-function is a BKP $\tau$-function in $\pb=(p_1,p_3,\dots)$ variables, and it is also a BKP $\tau$-function in
$\tb=(t_1,t_3,\dots)$ variables, see \cite{JM},\cite{KvdLbispec}.
\er

\bt
The $ \tau $-function  (\ref{tau+-}) $ \tau^\pm (\pb, \pb_1 | {\bf t}) $ that generates
the Hurwitz numbers 
is a two-component BKP multisoliton $ \tau$-function with respect to the time sets $ \tb $ and $ \pb $.
\et
For the proof, it suffices to represent  $\tau^\pm(\pb,\pb_1|{\bf t})$ in the form of an appropriate vacuum
expectation value.
Indeed, this 2-BKP (Veselov-Novikov) $\tau$-function $ {\tilde\tau}^{{\rm VN}\pm}(\pb,\tb)  $  is of the following
solitonic type
\be\label{BKPsolitons}
{\tilde\tau}^{{\rm VN}\pm}(\pb,\tb)=
\langle 0|\gamma^{(1)}(\tb)\gamma^{(2)}(\pb) g^\pm(0)
e^{\sqrt{-1}\sum_{i\ge 0}C_{ij}\phi^{(1)}(\zeta_i)\phi^{(2)}_j} |0\rangle
\ee
\be\label{BKPsolitons1}
=\sum_{\alpha,\beta\in\DP} 2^{-\tfrac12\bar{\ell}(\alpha)-\tfrac12\bar{\ell}(\beta)}C_{\alpha,\beta}
e^{\eta(\zeta_{\alpha_i},\tb) -\delta^\pm}\Delta^*(\zeta_\alpha)Q_\beta\{\pb\}
\ee
where the symbol $g^\pm(0)$ is given by (\ref{g}) where each $\phi_i$ is
replaced by $\phi_i^{(2)}$, and
where $ \eta(\zeta_{\alpha_i},\tb) $ and $\Delta^*(\zeta_\alpha)$ are given respectively
by (\ref{eta}) and by (\ref{Delta(zeta)}),
the set $ \{\zeta_i \} $ is a set of free parameters (soliton momenta), and a free chosen matrix
 $C$ defines an interaction between solitons (the net of lines in the $(t_1,t_3)$-plane in the tropical approximation
 of the $\tau$-function).
 The series (\ref{BKPsolitons1}) where
$
C_{\alpha,\beta}=\det \left(C\right)_{\alpha_i,\beta_j}
$
is obtained by direct calculation of the vacuum expectation value in (\ref{BKPsolitons}) taking into
account (\ref{multicorr}) and (\ref{Q+fermi}).
Then, choosing
$C_{j,0}=2^j\frac{1}{j!}\delta_{j,i},\,i>0$, $C_{j,0}=2\delta_{j,0}$ and
putting $\zeta_j=j$, one gets
${\tilde\tau}^{{\rm VN}\pm}(\pb,\tb)=\tau^\pm(\pb,\pb_1|{\bf t})$ as it follows from
(\ref{pm-hyp-tau}) and (\ref{dR}).

When all $t_i=0$ for $i>3$, this $\tau$-function generates the Hurwitz numbers
$ H^\pm\left(\Gamma_d^r,\Delta\right) $:
\be\label{tau+-p_1}
\tau^\pm(\pb,\pb_1|{\bf t})
=\sum_{d\ge 0}  e^{t_1d+t_3\left(d^2 -\tfrac 23 d\right)}
\sum_{\Delta\atop |\Delta|=d}
\sum_{b\ge 0} \frac{t_3^b}{b!}
H^\pm\left(\Gamma_d^b,\Delta\right)\pb_{\Delta}
\ee

Such a $\tau$-function describes ``the net of resonant solitons'' (it is either the net of regular
solitons or the net of singular solitons).\footnote{The best presentation of such nets of regular solitons
in the KP case is given in \cite{Kodama}. Singular resonant solitons were considered in \cite{O1983}.
Networks of singular solitons will be described elsewhere.}

Simirlarly to the previous case, one gets

\br\label{How-to-get-H(Gamma,Delta)}
To generate the Hurwitz numbers $H^\pm(\Gamma_d^b,\Delta)$, it is enough to consider the $N$-soliton
$\tau$-function with
the parameters $\zeta_i$ which fill the string $0,1,\dots,d$ where $d$ does not exceed the number of solitons $N$.
Such a $\tau$-function is holomorphic
 in the $(x,t)$-variables, and one can write
\be
\sum_{\Delta\atop |\Delta|=d} H^\pm\left(\Gamma^b_d,\Delta\right)\pb_\Delta = (2\pi)^{-1}\res_{t=0} t^{-b-1}
e^{-t\left(d^2 +\tfrac 23 d\right)}\int_{0}^{2\pi }dx e^{-ixd}  \tau_N^{{\rm VN},\pm}(ix,t)
\ee
The superscript ``-" denotes the singular soliton solution, while ``+" is related to the regular one.
\er

\subsection{Different fermionic expressions generating Hurwitz numbers \label{different}}

Besides formula (\ref{VEV1}) and the example discussed in (\ref{BKPsolitons}), there are many others
representations of the series (\ref{pm-hyp-tau}) in terms of the neutral fermions. These are different embeddings
of a given function of many variables into various families of $\tau$-functions.
Here are some more examples. In these examples,
$\eta(\alpha_i,\tb)$ and $\Delta^*(\zeta_\alpha)$ are given respectively by (\ref{eta}) and
by (\ref{Delta(zeta)}).

\bl\label{Lemma1} The multisoliton BKP $\tau$-function can be written as the following vacuum expectation value:
\be\label{tauA}
\tau_1(\tb)=
\langle 0|\gamma(\tb)e^{\sum_{i>j\ge 0} A_{ij}\phi(\zeta_i)\phi(\zeta_j)}|0\rangle
=\sum_{\alpha\in DP} 2^{-\tfrac 12 \bar\ell(\alpha)}{ A}_\alpha \Delta^*(\zeta_\alpha)\prod_{i=1}^{\ell(\alpha)}
e^{\eta(\zeta_{\alpha_i},\tb)}
\ee
where $ A_{\alpha}=\Pf \bar A$, and $\bar A$ is the  the skew symmetric
$\bar\ell(\alpha)\times \bar\ell(\alpha)$
submatrix of the matrix $A$ whose entries are selected by the parts of $\alpha=(\alpha_1,\dots,\alpha_k)$ as
\be
{\bar A}_{i,j} := A_{\alpha_i,\alpha_j},\quad i,j\le k
\ee
when $k$ is even, and
\be
{\bar A}_{i,j}: = \begin{cases}
                  A_{\alpha_i,\alpha_j},\quad i,j\le k \\
                  A_{\alpha_i,0},\,\quad i\le k,\,j=k+1
                 \end{cases}
\ee
when $k$ is odd.
For any choice of $A$, one gets the multisoliton
BKP $\tau$-function (which describes the net of resonant solitons with the momenta given by the set $\{\zeta_i\}$).
For $ \zeta_i=i $, one obtains the BKP multisoliton $\tau$-function with integer momenta.
\el

As an example, one can choose $A_{ij}=\frac 12 Q_{Ni,Nj}\{\pb\}F(i)F(j),\,j>0$ and $A_{i,0}= Q_{Ni,0}\{\pb\}F(i)$.
 Then, the $\tau$-function (\ref{tauA}) takes the form
\be\label{A-2}
\tau_1(\tb,\pb)=\sum_{\alpha\in\DP} 2^{-\ell(\alpha)}  \Delta^*(\zeta_\alpha) Q_{N\alpha}\{\pb\})
\prod_{i=1}^{\ell(\alpha)}e^{\eta(\zeta_{\alpha_i},\tb)} F(\alpha_i)
\ee

 The right hand side of (\ref{A-2}) where $F(i)=\pm 1$ can be equated to the $\tau$-function (\ref{BKPsolitons}) if we choose
 $C_{i,j} \sim\delta_{Ni,j}$ in (\ref{BKPsolitons1}).

If we put $ p_k = \delta_ {k, 1} $ and $F(i)=\pm\frac{(Ni)!}{i!}$,  we get
KdV $ \tau $-function (\ref{KdV-sol}).

\bl\label{Lemma2} Another  example of the BKP $\tau$-function ($A_\alpha$ is the same as in the previous Lemma) is given by
\be\label{tauB}
\tau_2^{\rm}(\pb)=
\langle 0|\gamma(\pb)e^{\sum_{i>j\ge 0} A_{ij}\phi_i\phi_j}|0\rangle
=\sum_{\alpha} 2^{-\tfrac 12 {\bar\ell}(\alpha)}A_\alpha Q_\alpha\{\pb\}
\ee
\el
If we choose $A$ to be the same as in example (\ref{A-2}), the $\tau$-function (\ref{tauB})
takes the form
\be\label{tauB'}
\tau_2^{\rm}(\pb,\pb^*) =
\sum_{\alpha\in\DP}2^{- {\ell}(\alpha)}
 Q_\alpha\{\pb\}Q_{N\alpha}\{\pb^*\}\prod_{i=1}^{\ell(\alpha)} F(\alpha_i)
\ee
which is of the form considered in sec.\ref{tau-from-characters} below. For $N=1$, (\ref{tauB'}) is the
hypergeometric $\tau$-function given by
(\ref{pm-hyp-tau}) if we choose $F(j)=\pm\exp \sum_m j^m t_m$.

\bl\label{Lemma3} An example of the 2-component BKP (2-BKP) $\tau$-function is given by
\be\label{tauC}
\tau_3(\pb^{(1)},\pb^{(2)})=
\langle 0|\gamma^{(1)}(\pb^{(1)})\gamma^{(2)}(\pb^{(2)})e^{\sqrt{-1}\sum_{i,j\ge 0} C_{ij}\phi^{(1)}_i\phi^{(2)}_j}|0\rangle
=\sum_{\alpha,\beta\in \DP} 2^{-\tfrac 12 {\bar\ell}(\alpha)-\tfrac 12 {\bar\ell}(\beta) }C_{\alpha,\beta}
Q_\alpha\{\pb^{(1)}\}Q_\beta\{\pb^{(2)}\}\nn\\
\ee
where $C_{\alpha,\beta} =\det \left( C \right)_{\alpha_i,\beta_j}$.
\el
For instance, one can choose $C_{ij}=F(j)\delta_{i,Nj},\,j>0$, $C_{i,0}=2\delta_{i,0}$ in order to get once again the series
at the right hand side of (\ref{tauB'}).

\bl\label{Lemma4} Another example of the 2-BKP $\tau$-function is given by
\be\label{tauC2}
\tau_4(\pb^{(1)},\pb^{(2)})=
\langle 0|\gamma^{(1)}(\pb^{(1)})\gamma^{(2)}(\pb^{(2)})
e^{\sqrt{-1}\sum_{i,j\ge 0} C_{ij}\phi^{(1)}(\zeta_i)\phi^{(2)}(\zeta_j)}|0\rangle
\ee
\be
=\sum_{\alpha,\beta\in \DP} 2^{-\tfrac 12 {\bar\ell}(\alpha)-\tfrac 12 {\bar\ell}(\beta) }C_{\alpha,\beta}
 \prod_{i=1}^{\ell(\alpha)}e^{\eta(\zeta_{\alpha_i},\pb^{(1)})}\prod_{i=1}^{\ell(\alpha)}
 e^{\eta(\zeta_{\beta_i},\pb^{(2)})}
 \Delta^*_\alpha(\zeta^{(1)})
 \Delta^*_\beta(\zeta^{(2)})
\nn\\
\ee
\el
To get the KdV multisoliton $\tau$-function, it is enough to take $\zeta_i^{(1)}=\zeta_i^{(2)},\,i\ge 0$. To get the $\tau$-function
(\ref{KdV-sol}), one takes $\zeta_i=i$ and $C_{ij}= 2^{2j}(j!)^{-2} \delta_{i,j},\,j>0 $,
$C_{i,0}=C_{0,i}=2\delta_{i,0} $.

The examples of the KdV and Veselov-Novikov soliton $\tau$-functions considered above can be presented as a specification of
$\tau_1,\dots,\tau_4$.

\br
In all cases, the sums in the exponentials can be replaced by integrals with appropriate measures
(to get specializations of the hypergeometric $\tau$-functions discussed in sec.\ref{KdV}, one needs to choose
the $\Gamma$-function and the contour going around the real semi-axis in the complex plane).
\er
Any particular fermionic representation is convenient for writing down the string equations, which are stationary points for
special combinations of $ W^B_\infty $ symmetries discussed in sec.\ref{W}.

\subsection{Matrix model $\tau$-functions as fermion averages \label{MM}}

The partition functions of the matrix models in external field considered in the Introduction, (\ref{Kon}) and
(\ref{BGW}),
which are hypergeometric BKP $\tau$-functions can be considered as particular cases of described fermion averages.

The $\tau$-function of the BGW model in (\ref{BGW}) can be
presented as examples considered in sec.\ref{different} by a proper specification
of parameters:
\begin{proposition}

\be\label{Aleksandrov}
\tau_{BGW}(\pb) =
\sum_{\alpha\in\DP} 2^{-\ell(\alpha)}
Q_\alpha\{\pb\}Q_{\alpha}\{\delta_{k,1}\}\prod_{i=1}^{\ell(\alpha)}\left(\frac{(2\alpha_i)!}{\alpha_i!}\right)^2
\ee
$$
=\langle 0|\gamma(\pb) e^{\sum_{n>0} \log\left( \frac{(2n)!}{n!}\right)^2(-1)^n \phi_n\phi_{-n}}
\gamma\{\delta_{k,1}\}|0\rangle =
$$
$$
\langle 0|\gamma(\pb)e^{\sum_{i>j\ge 0} \frac 12
\left(\frac{(2i)!(2j)!}{i!j!}\right)^2 {\tilde Q}_{i,j}\phi_i\phi_j  }|0\rangle=
\langle 0|\gamma^{(1)}(\pb)\gamma^{(2)}\{\delta_{k,1}\}
e^{\sqrt{-1}\sum_{i\ge 0} C_i\left(\frac{(2i)!}{i!}\right)^2\phi^{(1)}_i\phi^{(2)}_{i}}|0\rangle
$$
where we used (\ref{dR}). Here
${\tilde Q}_{i,j}={Q}_{i,j}\{\delta_{k,1}\}= 2^{i+j}\frac{1}{i!j!}\frac{i-j}{i+j}$;
${\tilde Q}_{i,0}=2{Q}_{i,0}\{\delta_{k,1}\}=2^{i+1}\frac{1}{i!}$ and $C_i=1,\,i>0;\, C_0=2$

\end{proposition}

Similarly, the $\tau$-function of the Kontsevich model (\ref{Kon}) can be presented as as examples considered in sec.\ref{different}:
\begin{proposition}
The series (\ref{Kon}) can be treated both as the BKP and as the 2-BKP $\tau$-function presented as a fermion average
\be\label{tauC-Kon}
\tau_{K_3}\{\pb\}=
\sum_{\alpha\in\DP} 2^{-\ell(\alpha)} Q_\alpha\{\pb\}Q_{2\alpha}\{\delta_{k,3}\}\prod_{i=1}^{\ell(\alpha)}
\frac{(2\alpha_i)!}{\alpha_i!}=
\ee
$$
=\langle 0|\gamma(\pb)e^{\sum_{i>j\ge 0} \frac 12
\frac{(2i)!(2j)!}{i!j!} {\tilde Q}_{2i,2j}\{\delta_{k,3}\}\phi_i\phi_j  }|0\rangle
=
\langle 0|\gamma^{(1)}(\pb)\gamma^{(2)}\{\delta_{k,3}\}
e^{\sqrt{-1}\sum_{i\ge 0} C_i\frac{(2i)!}{i!}\phi^{(1)}_i\phi^{(2)}_{2i}}|0\rangle
$$
where ${\tilde Q}_{2i,2j}={Q}_{2i,2j}\{\delta_{k,3}\}$ and ${\tilde Q}_{2i,0}=2{Q}_{2i,0}\{\delta_{k,3}\}$
and where $C_i=1,\,i>0$ and $C_0=2$.
\end{proposition}

\subsection{On character expansion for KdV $\tau$-functions and BKP}

Since the bilinear Pl\"ucker relations (\ref{Plucker}) does not contain terms $c_\alpha^2$ for any diagram $\alpha$,
one can put all the coefficients $c_\alpha=0$ but one. Hence, any individual Schur function solves the Pl\"ucker
relations, and is a KP $\tau$-function. For a similar reason, any individual $Q$ Schur function is a $\tau$-function
of the BKP hierarchy.

If one applies the Schur function expansion, (\ref{Sex}) to the KdV hierarchy, which is a KP reduction, one immediately
realizes that an individual Schur function (i.e. a single non-vanishing $c_\alpha$) solves the hierarchy only for the
diagrams $[\ldots, 4,3,2,1]$. This is not surprising, since the KdV $\tau$-function does not depend on even time
vatriables $p_{2k}$, and so do only these Schur functions:
${\rm Schur}_{[1]} = p_1$ or ${\rm Schur}_{[2,1]} = \frac{1}{3}(-p_3+p_1^3)$, etc.

On the other hand, the KdV solutions simultaneously solve the BKP hierarchy, and one could expect that a natural c
haracter expansion of the KdV hierarchy is in terms of the $Q$ Schur functions (\ref{Qex}).
Of course, this is formally true, because the $Q$ Schur functions form a complete basis in the space of functions
of odd times $p_{2k+1}$, but the restrictions on the coefficients $c_\alpha$ substituting the Pl\"ucker relations
in  KP and BKP cases, is more involved. In particular, most individual $Q$-functions are not the KdV $\tau$-functions:
exceptions are provided just by $Q_{\ldots, 4,3,2,1}$, which coincide with peculiar ordinary Schur functions,
depending on odd times only as discussed above.

Note that the reduction BKP $\to$ KdV equation has the simplest form when one uses the 2-BKP approach and imposes the
condition
\be
\left(\frac{\partial}{\partial p^{(1)}_1}-\frac{\partial}{\partial p^{(2)}_1}\right)\tau(\pb^{(1)},\pb^{(2)})=0
\ee
and the KdV higher times can be identified with $ \pb^{(1)}+\pb^{(2)}$.
It can be also formulated in terms of the Lax operator for the Veselov-Novikov equation, which is
a 2D Schr\"odinger operator (with a reduction 2D Sch\"odinger $\to$ 1D Schr\"odinger operator the latter being
the Lax operator for the KdV equation). Another way is to present the KdV hierarchy as the so-called rational
reduction of the BKP hierarchy.
For some other aspects of relation between the KdV and BKP hierarchies, see \cite{DJKM1,JM,DJKM2,DJKM,Alex2}.

\section{Factorization on special loci}

\subsection{Specialization at $p_k=\delta_{k,r}$}

The locus $p_k=\delta_{k,1}$ is well known to play a big role in representation theory and character calculus. For
instance, in the case of Schur polynomial $S_R$, this locus is associated with the dimension $|
R|!S_{R}\{\delta_{k,1}\}$ of representation $R$ of the permutation group $S_{|R|}$ and reflects the Schur-Weyl
duality \cite{Fulton2}. In this case, this quantity has a special notation, $d_R:= S_{R}\{\delta_{k,1}\}$.  Somewhat
unexpectedly superintegrability relations in the Gaussian matrix models are sensitive also to the values at other
loci $p_k=\delta_{k,r}$ \cite{MMq}. This gets a natural explanation in the study of monomial matrix models, where
the Cauchy identity states
\be
e^{-\frac{1}{r}\,\Tr X^r} = \exp\left(-\sum_k \frac{1}{k}\,\Tr X^k \cdot \delta_{k,r}\right) =
\sum_R (-1)^{|R|}\,S_{R^\vee}\{\Tr X^k\} \cdot S_R\{\delta_{k,r}\}
\ee
This makes the study of character values at these delta-loci very important. These values are known to be
distinguishably factorizable, and very recently A.Alexandrov \cite{Alex} conjectured an explicit formula for
the ratio of two $Q$ Schur functions, $Q_R/Q_{2R}$ at $p_k=\delta_{k,3}$. In this section, we describe a general
and explicit factorization formula for $Q_R/Q_{NR}$ at any $N$ and $p_k=\delta_{k,r}$ (hence, $r$ is odd), and even
more general factorization formula for $Q_R$ at only $p_{k,rj}$ non-vanishing ($r$ and $j$ odd). We prove these
formulas in Appendices A and B.


The relation important for us is
\be\label{Nr}
Q_{\alpha}\{\delta_{k,r}\}= Q_{N\alpha}\{\delta_{k,r}\}\prod_{i=1}^{\ell(\alpha)} F(\alpha_i)
\ee
with coprime $N$ and $r$ and with some function $F(\alpha_i)$ that we describe below.

In order to explain (\ref{Nr}), we list a set of facts about specializations of the symmetric functions at
$p_k=\delta_{k,r}$.

\begin{itemize}
\item For the Schur functions, the values at $p_k^{(r)}=\delta_{k,r}$ are equal to
\be
S_R\{\delta_{k,r}\}=\delta_r(R)\prod_{x\in R}{1\over [h_x]_{0,r}}
\ee
where $h_x$ is the hook length and $[n]_{p,r}$ is defined to be $n$ when $n$ is equal to $p$ mod $r$, and to be
1 otherwise. $\delta_r(R)$ is defined in \cite[Eq.(3.26)]{Pop}:
\be
\delta_r(R)=\left\{
\begin{array}{cl}
(-1)^{|R|/r}\prod_{x\in R}(-1)^{[c_x/r]+[h_x/r]}&\hbox{ if the $r$-core of $R$ is trivial}\cr\cr
0&\hbox{otherwise}
\end{array}
\right.
\ee
where $c_x$ is the content of the box in $R$.
\item Similarly, for the $Q$ Schur functions, the values at $p_k^*=\delta_{k,r}/2$ are equal to
\be
Q_\alpha\{\delta_{k,r}/2\}=\tilde\delta_r(\alpha)\prod_{x\in S(\alpha)}{1\over [h^{(d)}_x]_{0,r}}
\ee
where $h^{(d)}_x$ is the hook length in the doubled Young diagram, which is defined to be
$d(\alpha):=(\alpha_1,\alpha_2,\dots|\alpha_1-1,\alpha_2-1,
\ldots)$ in the Fr\"obenius notation \cite{Mac}, while $S(\alpha)$ denotes here the shifted Young diagram \cite{Mac}, which is the part of $d(\alpha)$ lying above the main diagonal. The sign factor $\tilde\delta_r(\alpha)$ vanishes if the $r$-core of $d(\alpha)$ is
non-trivial.

\item Suppose we are given a strict partition $\alpha$ such that the size of $\alpha$ is divisible by $r$, and
$\tilde\delta_r(\alpha)\ne 0$ (this simultaneously implies that $\tilde\delta_r(N\alpha)\ne 0$), i.e. the $r$-core of
$d(\alpha)$ is trivial. Suppose also that $N$ and $r$ are coprime ($r$ is certainly odd). Then, the following formula is correct:
\be\label{f}\boxed{
{Q_{\alpha}\{{r\over 2}\cdot\delta_{k,r}\}\over Q_{N\alpha}\{{r\over 2}\cdot\delta_{k,r}\}}= \prod_{i=1}^{\ell(\alpha)}
(-1)^{\rho_{_{N,r}}(\alpha_i)}\cdot N^{ \{\alpha_i/r\} }\cdot  {[N\alpha_i/r]! \over [\alpha_i/r]!}}
\ee
where $\{\ldots\}$ at the r.h.s. denotes the fractional part of a number, and $[\ldots]$ denotes the integer part.
\end{itemize}

We derive formula (\ref{f}) in Appendix A, here just point out that the integer-valued function $\rho_{N,r}(x)$
depends only on $(x)_r$ (the value of $x$ mod $r$), and on $(Nx)_r$ (the value of $Nx$ mod $r$). Manifestly,
$\rho_{_{N,r}}(0)=0$, and all other $x_r$ enters the product in pairs $\rho_{_{N,r}}(k)+\rho_{_{N,r}}(r-k)$ so
that there is no difference which sign to choose for an individual $(-1)^{\rho_{_{N,r}}(x)}$ in the pair. For the
sake of definiteness, let us choose $(-1)^{\rho_{_{N,r}}(k)}=1$ with $r\ge k>r/2$. Then one has
\be\label{sign}
\rho_{_{N,r}}(x)=\left\{
\begin{array}{cl}
(Nx)_r-(x)_r &\hbox{for }0<(x)_r<r/2\cr
\cr
0& \hbox{otherwise}
\end{array}\right.
\ee

In fact, there is a more fundamental factorization formula for $p_k$ non-vanishing only at $k$ divisible by $r$,
(\ref{Q-S}), and formula (\ref{f}) is its straightforward corollary in a particular case. We discuss this in the
next subsection.

\subsection{Basic factorization formula}

Now we consider a more general factorization formula in the case, when an infinite set of times $t_{kr}$ is
non-vanishing. That is, we introduce a set of times  $\pb[r]:=\left(p_1[r],p_3[r],\dots\right)$ such that
\be \label{t[r]-t}
\frac 1k p_k[r]=\frac 1j p_j \delta_{k,jr},\quad {\rm both}\quad r,j\,\,{\rm odd}
\ee
Let us consider a strict partition $\alpha$, and produce $r$ new strict partitions $\mu$, $a^c$, $b^c$ made of
$[\alpha_i/r]$. Parting into these $r-1$ partitions depends on the value $(\alpha_i)_r=x$: the parts with $x=0$
get to partition $\mu$, the parts with $0<x\le (r-1)/2$ get to partitions $a^c$, $c=x$, and those with $(r-1)/2<x<r $
get to partitions $b^c$, $c=r-x$. Thus, the partition $a^c$ at each color $c$ has an associated partition $b^c$.

Thus, the parts of $\alpha$ are parted into three groups:
 \begin{itemize}
  \item parts $r\mu$ that are divisible by $r$
  \item parts presented as $ra^c+c$ where $c=1,\dots, \frac 12 (r-1)$
  \item parts presented as $r(b^c+1)-c$ where $c=1,\dots, \frac 12 (r-1)$
 \end{itemize}
Suppose that $|\alpha|$ is divisible by $r$ and that the length of partitions $a^c$ coincides with those of $b^c$ (otherwise, $Q_{\alpha}\{p_k[r]\}=0$).  We denote this length through
$\kappa^c$. Then, there is a beautiful factorization
formula (which we derive in Appendix B)
\be\label{Q-S}
\boxed{
Q_{\alpha}\{p_k[r]\}=(-1)^\omega\ 2^{-\tfrac12{{\bar\ell}(\mu)}}Q_{\mu}\{p_k\}\prod_{c=1}^{\tfrac 12 (r-1)}
\!\!\!S_{(a^c|b^c)}\{2p_k'\}\cdot
\prod_{i=1}^{\kappa^c}(-1)^{a_i^c+b_i^c+c}
}
\ee
where $p_k':=(p_1,0,p_3,0,p_5,\ldots)$, $S_{(a^c|b^c)}$ is the ordinary Schur function in the Fr\"obenius (hook)
notation, and $\omega$ depends on the order of embedded parts which belong to one of the three groups.
Basically, it is not important for our purposes because we will be interested in rescaling
of lengths of the parts $\alpha_i \to N\alpha_i$ which keeps the order, and we get the same $\omega$.

For example, for $r=3$ and a partition, say, $\alpha=[6,5,4,3,2,1]$, $\mu=[2,1]$, $a^1=[1,0]$, $b^1=[1,0]$ so that
eq.(\ref{Q-S}) states:
\be
\!\!\!\!\!\!\!
Q_{[6,5,4,3,2,1]}\{0,0,3p_1,0,0,0,0,0,3p_3,\ldots\} \sim Q_{[2,1]}\{p_1,p_3\} \cdot
\underbrace{S_{(1,0|1,0)}}_{S_{[2,2]}}\{2p_1,0,2p_3,0\}
\ee
Similarly, for $r=3$ and a partition $\alpha=[15,7,6,5,2,1]$, $\mu=[5,2]$, $a^1=[2,0]$, $b^1=[1,0]$
so that eq.(\ref{Q-S}) states:
\be
\!\!\!\!\!\!\!
Q_{[15,7,6,5,2,1]}\{0,0,3p_1,0,0,0,0,0,3p_3,\ldots\} \sim Q_{[5,2]}\{p_1,p_3,p_5,p_7\} \cdot
\underbrace{S_{(2,0|1,0)}}_{S_{[3,2]}}\{2p_1,0,2p_3,0,p_5\}
\ee

The basic factorization formula (\ref{Q-S}) immediately leads to formula (\ref{f}), see Appendix A for details.

\section{Hypergeometric $\tau$-functions entirely made from characters \label{tau-from-characters} }

In this section, we describe an important construction of particular hypergeometric $\tau$-functions from the
ratios of characters at the special loci $p_k = \delta_{k,r}$ described in the previous section. It is actually
applicable in a more general context, but we present it in the case of the $Q$ Schur functions and the BKP hierarchy.


The relations (\ref{Nr}), (\ref{f}) implies that the bilinear combination (\ref{1}), which is a BKP $\tau$-function
can be specified in the form
\be\label{sp}
\tau_{BKP}\{p,\bar p\} = \sum_{\alpha\in\DP} Q_\alpha\{p\}Q_{N\alpha}\{\delta_{k,r}\} \cdot \prod_{i=1}^{\ell(\alpha)}
f(\alpha_i)
\ee
Indeed, the restrictions for $\alpha$ in (\ref{f}) implies that $Q_{N\alpha}\ne 0$ and, hence, are satisfied, since
otherwise
$\alpha$ does not contribute to the sum (\ref{sp}).

In its turn, this means that not only the partition function of the Kontsevich model, \cite{MMq}
\be
\tau_{K_3}\{p_k\} = \sum_{\alpha\in\DP} {1\over 2^{\ell(\alpha)}}\cdot
{Q_\alpha\{p_k\}Q_{\alpha}\{\delta_{k,1}\}Q_{2\alpha}\{\delta_{k,3}\}\over Q_{2\alpha}\{\delta_{k,1}\}}
\ee
which is of form (\ref{sp}) with $N=2$, $r=3$ \cite{Alex}, is a BKP $\tau$-function  with the weight function given by
\be
\prod_{i=1}^{\ell(\alpha)} f(\alpha_i)={Q_{\alpha}\{\delta_{k,3}\}
\over Q_{2\alpha}\{\delta_{k,3}\}}{Q_{\alpha}\{\delta_{k,1}\}\over Q_{2\alpha}\{\delta_{k,1}\}}
\ee
but a more general combination with arbitrary coprime $N$ and $r$ is still a BKP $\tau$-function.

From (\ref{f}), it also follows that one can choose it as a product of various ratios
${Q_{\alpha}\{\delta_{k,r}\}\over Q_{N\alpha}\{\delta_{k,r}\}}$ still preserving the BKP $\tau$-function:
\be\boxed{
\tau_{BKP}\{p,\bar p\} =\sum_{\alpha\in\DP} Q_\alpha\{p\}Q_{\alpha}\{\delta_{k,r}\}\prod_i
{Q_{\alpha}\{\delta_{k,r_i}\}\over Q_{N_i\alpha}\{\delta_{k,r_i}\}}\prod_j{Q_{N_j\alpha}\{\delta_{k,r_j}\}
\over Q_{\alpha}\{\delta_{k,r_j}\}}}
\ee
with arbitrary sets of (coprime) $N_i,\ r_i,$ and $N_j,\ r_j$, however, one has to deal carefully with vanishing of $Q_{\alpha}\{\delta_{k,r_i}\}$ (either choosing proper sets of $r_i$'s, or continuing formula (\ref{f}) to all sets of $\alpha_i$ by definition).

This means that one can construct non-trivial
hypergeometric BKP $\tau$-functions completely in terms of characters. The cubic Kontsevich and BGW partition
functions belong to this class, which once again emphasizes {\bf the intimate connection between  the hypergeometric
and matrix model partition functions}, i.e. between hypergeometricity and string equations.

Since the Virasoro algebra acts on the $Q$ Schur functions in a very simple way \cite{Aok}:
\be
\hat L_n Q_\alpha \left\{\frac{p_k}{\sqrt{2}}\right\}= \sum_{i=1}^{\ell(\alpha)}
  (-)^{\nu_i}(\alpha_i- n)\
 Q_{\alpha-2n\epsilon_i}\left\{\frac{p_k}{\sqrt{2}}\right\},\nn\\
\hat L_n:=\sum_{k\in\mathbb{Z}^+_{odd}}(k+n)p_k{\partial\over\partial p_{k+n}}
+{1\over 2}\sum_{a,b\in\mathbb{Z}^+_{odd}}^{a+b=2n}ab{\partial^2\over\partial p_a\partial p_b}
\ee
where $\alpha-2k\epsilon_i$ denotes the shift of $\alpha_i\to \alpha_i-2k$. This can make it shorter than some other lines
and thus imply reordering of lines in the diagram to put them back into decreasing order,
then $\nu_i(\alpha,b)$ is the number of lines, which the $i$-th one needs to jump over. It is possible to check that (\ref{Kon})
and (\ref{BGW}) satisfy the Virasoro constraints  (after an appropriate rescaling of time variables). This illustrates
the general claim of \cite{MM} that superintegrability of matrix models, which underlies these expressions implies
both the ordinary integrability and the Virasoro constraints: the two basic properties of matrix model partition
functions \cite{UFN3}. From the very beginning, it was clear that they are intimately related \cite{FKN}, but a
nature of this relation remained obscure. Now we understand that they are just two different corollaries of a more
fundamental superintegrability feature $<character> = character$.

\section{Conclusion}

In this paper, we made a brief review of the details
behind the formalism of $Q$ Schur functions, which can be relevant for
deeper investigation of the spin Hurwitz partition functions.
We put a special emphasize on the ``hypergeometric" $\tau$-functions.
They are made from Casimir exponentials and give rise
to peculiarly-factorized coefficients in the character expansions,
which is typically associated with (generalized) hypergeometric series, hence the name.
There is a mounting evidence that matrix model $\tau$-functions
(i.e. those which satisfy additional string equations)
belong to this class.
Until recently, this was not so obvious, because important
matrix models were not brought to this form,
but recently the reason has been found:
the $\tau$-functions for the Kontsevich and BGW models
are expanded in the $Q$ Schur functions rather than in the ordinary Schur functions.
A posteriori, this is rather obvious because they satisfy
KdV rather than the generic KP/Toda hierarchy.
Moreover, now it is clear that the generalized Kontsevich model
is likewise expressed through appropriately generalized $Q$ Schur functions
\cite{MMgkm}.
In this paper, we, however, concentrated on the standard $Q$ Schur functions
and perspectives of spin Hurwitz studies.

\paragraph{Acknowledgements.} \small{We are grateful to A. Alexandrov for various discussions
and for sharing his results prior to publication. A. O. thanks J. van de Leur and J. Harnad for numerous
discussions on topics related to the BKP hierarchy.
The publication was partly prepared by Sergey Natanzon within the framework of the Academic Fund Program
at the National Research University Higher School of Economics (HSE)
in 2020-2021 (grant 20-01-009) and by the Russian Academic Excellence Project “5-100”.
A. O. was supported  by RFBR grant 18-01-00273a.
Our work is also supported in part by the grant of the
Foundation for the Advancement of Theoretical Physics ``BASIS" (A.Mir.),
by  RFBR grants 19-01-00680 (A.Mir.) and 19-02-00815 (A.Mor.),
by joint grants 19-51-50008-YaF-a (A.Mir.), 19-51-53014-GFEN-a (A.Mir., A.Mor.), 21-52-52004-MNT-a (A.Mir., A.Mor.).
The work was also partly funded by RFBR and NSFB according
to the research project 19-51-18006 (A.Mir., A.Mor.).}
\bigskip
\bigskip

\appendix

\section{Derivation of (\ref{f}) from (\ref{Q-S})}

The basic factorization formula (\ref{Q-S}) allows us to obtain formula (\ref{f}) immediately with the following logic. First of all, the basic factorization formula (\ref{Q-S}) expresses the $Q$ Schur functions, $Q_R$ at $p_k[r]$ through those at $p_k$, these two being related by (\ref{t[r]-t}).
In particular, under the choice $p_k[r]=r\delta_{k,r}$, it expresses the $Q$ Schur functions at $\delta_{k,r}$ through those at $\delta_{k,1}$.
Hence, it remains to evaluate the $Q$ Schur functions at $p_k=\delta_{k,1}$, and then insert the obtained results into (\ref{Q-S}) in order to obtain finally (\ref{A9}).
This is what is done in this Appendix in three steps.
The factorization formula itself is proved in Appendix B.

\paragraph{Step I. (\ref{Q-S}) explicitly at $p_k=r\delta_{k,r}/2$.}

In order to obtain formula (\ref{f}), we evaluate (\ref{Q-S}) at $p_k=r\delta_{k,r}/2$. Using (\ref{dR}), we get for
individual factors in (\ref{Q-S}) in this case:
\be\label{a1}
Q_{\mu}\left\{{1\over 2}\cdot\delta_{k,1} \right\}=\prod_{i=1}^{\kappa^0}{1
\over \mu_i!}\prod_{i<j}\frac{\mu_i-\mu_j}{\mu_i+\mu_j}
\ee
\be\label{a2}
S_{(a^c|b^c)}\{\delta_{k,1} \}=\frac{1}{\prod_{i=1}^{\kappa^c}a^c_i!b^c_i!}
\frac{\prod_{i<j}(a^c_i-a^c_j)(b^c_i-b^c_j)}{\prod_{i,j}(a^c_i+b^c_j+1)}
\ee

\paragraph{Step II. (\ref{Q-S}) for $N\alpha$.}

Now we evaluate (\ref{Q-S}) for $N\alpha$. To this end, we notice that $N\alpha$ consists of the following parts:
\begin{itemize}
  \item parts $N r\mu$ that are still divisible by $r$
  \item parts presented as $Nra^c+Nc$ where $Nc=N,\dots, \frac 12 N(r-1)$
  \item parts presented as $Nr(b^c+1)-Nc$ where $Nc=N,\dots, \frac 12 N(r-1)$
 \end{itemize}
For each $c$ there exists $p_c$ and $c_N$
\be\label{c-c_N}
Nc=rp_c+c_N,\quad c_N < r
\ee
Let
\be\label{A}
N<r,\ \ \ \ \ \ \ \ \
N\,{\rm and}\,r\,\hbox{are coprime}
\ee
Equation (\ref{c-c_N}) maps each $c$ to a certain $c_N$.
If $ c $ and $ c'$ are different, then $ c_N (c) $ can not coincide with
either $r-c_N(c')$ or $c_N(c')$ .
Indeed,  suppose it is not correct, and one has the same $c_N$ for two different $c$ ($c$ and $c'$):
$$
N(c+c')=r(p_c+p'_c+r)\,\Rightarrow\,c+c'=\frac{r}{N}(p_c+p_c'+1)
$$
However, this is impossible because of (\ref{A}) and because both $c$ and $c'$ are less than $\frac 12 r$.
Similarly impossible is
$$
N(c-c')=r(p_c-p_c')\,\Rightarrow\,c-c'=\frac{r}{N}(p_c-p_c')
$$
Notice that the partition associated by $c$ is also associated by $c_N$:
\be
(a^c,b^c)=\left(ra^c_i +c\,,\,r(b^c_j+1) -c\right)\,\to\,
\left(Nra^c_i +Nc\,,\,Nr(b^c_j+1) -Nc\right)=\nn\\
=\left(r(Na_i^{c}+p_c) +c_N\,,\,r(Nb_j^{c}-p_c) -c_N\right)
\ee
Thus, one finally obtains
\be
Na=r(Na_i +p_i)+c_N,\quad Nb=r(N(b_i+1) - p_i) -c_N
\ee
Therefore, one gets
\be\label{Na1}
S_{(Na^c|Nb^c)}\{\delta_{k,1} \}=
\frac{  N^{-\kappa^c} }{\prod_{i=1}^{\kappa^c} (Na^c_i+p_i)!(Nb^c_i -p_i-1) !}
\frac{\prod_{i<j}(a^c_i-a^c_j)(b^c_i-b^c_j)}{\prod_{i.j=1}^{\kappa^c}(a^c_i+b^c_j+1)}
\ee
and
\be\label{Na2}
Q_{N\mu}\left\{{1\over 2}\cdot\delta_{k,1} \right\}=\frac{1}{\prod_{i=1}^{k} (N\mu_i)!}
\prod_{i<j}\frac{\mu_i-\mu_j}{\mu_i+\mu_j}
\ee

\paragraph{Step III. Evaluating the ratio (\ref{f}).}

Now, using (\ref{a1})-(\ref{a2}) and (\ref{Na1})-(\ref{Na2}), we are ready to evaluate the ratio of the $Q$
Schur functions in (\ref{f}):
\be\label{A9}
{Q_{\alpha}\left\{{r\over 2}\cdot\delta_{k,r}\right\}\over Q_{N\alpha}\left\{{r\over 2}\cdot\delta_{k,r}\right\}}=
(-1)^{g}
\left( \prod_{i=1}^{\ell(\mu)}\frac{(N\mu_i)!}{\mu_i!} \right)
\prod_{c=1}^{\frac12(r-1)}N^{\kappa^c}\prod_{i=1}^{\kappa^c}\frac{(Na^c_i+p_c)!(Nb^c_i-p_c-1)!   }{a^c_i!b^c_i !}
=(-1)^g N^{\tfrac 12 v}\prod_{i=1}^{\ell(\alpha)}\frac{[N\alpha_i/r]!}{[\alpha_i/r]!}\nn\\
\ee
where $v$ is the number of parts of $\alpha$ that are not divisible by $r$, and
$g$ originates from the sign factors at the r.h.s. of (\ref{Q-S}) and is equal to
\be
g=\sum_{c=1}^{\frac12(r-1)}\kappa^c(c-c_N(c))
\ee
where $c_N(c)$ is given by (\ref{c-c_N}). Since this sign factor coincides with (\ref{sign}) and
$v=2\sum_{i=1}^{\ell(\alpha)}\{\alpha_i/r\}$, we finally come to formula (\ref{f}).

\section{Factorization formula (\ref{Q-S}) from fermion calculus}

Here we prove (\ref{Q-S}) using the fermion average representation for the $Q$ Schur functions (\ref{Q+fermi}).
To this end, we consider the fermionic representation for the $Q$ Schur function (\ref{Q+fermi})
\be\label{Q_R}
Q_{\alpha}\{p_k[r]\}=2^{\bar\ell(\alpha)}\langle 0|\gamma(\pb[r])\phi_{\alpha_1}\cdots \phi_{\alpha_l} |0\rangle=
2^{\bar\ell(\alpha)}\langle 0|\phi_{\alpha_1}(\pb[r])\cdots \phi_{\alpha_l}(\pb[r]) |0\rangle
\ee
$$
\phi_k(\pb[r]):=\gamma(\pb[r])\cdot\phi_{\alpha_1}\cdot\gamma(\pb[r])^{-1}
$$
Using the canonical anticommutation relation (\ref{neutral-canonical}), we have
 \be
 [J_m , \phi_i] = \phi_{i-m},\quad m\,{\rm odd}
 \ee
 and
 \be\label{dynamics}
 \phi_j(\pb[r]):=e^{\sum_{m>0,{\rm odd}} \frac{2}{nr}J_{mr}p_{mr}[r]}\phi_j
 e^{-\sum_{m>0,{\rm odd}} \frac{2}{nr} J_{mr}t_{mr}[r]}=
 \sum_{m\ge 0}\phi_{j-mr}h_m\{2p_k'\}
 \ee
 where $h_i$ are complete symmetric functions restricted on the set of odd labeled times:
\be\label{h}
e^{\sum_{n>0, {\rm odd}} \frac 2n p_n z^n} =\sum_{n\ge 0} z^n h_n\{2p_k'\}, \quad h_n\{2p_k'\}:= S_{(n)}\{2p_k'\}
\ee
Let us note that the exponential (\ref{h}) is also a generating function for the elementary projective
Schur functions $Q_{n,0}$:
\be\label{h=q}
h_n\{2p_k'\}=Q_{(n,0)}\{p_k\}
\ee

For evaluation of the averages at the r.h.s of (\ref{Q_R}), we use the Wick theorem. To this end,
we consider the pairwise average:
 \be\label{pair-corr}
-\langle 0| \phi_{a}(\pb[r])\phi_{b}(\pb[r]) |0\rangle =
\langle 0| \phi_{b}(\pb[r])\phi_{a}(\pb[r]) |0\rangle
 =\langle 0|  \sum_{m\ge 0}\phi_{b-mr}h_m\{2p_k'\} \sum_{n\ge 0}\phi_{a-nr}h_n\{2p_k'\} |0\rangle
 \ee
where, according to the canonical pairing (\ref{b-pairing}), contribute only the terms with
 \be\label{condition}
a - nr + b-mr =0
 \ee
This implies distributing the parts of the strict partition into three groups $\mu$, $a^c$ and $b^c$ as above.

\br
 One may say that there exist a part $\alpha_x$ such that $a^c_i +p_c=[\alpha_x /r]$ and $c=(\alpha_x)_r$, and a
 part $\alpha_y$, with $[\alpha_y/r]=b^c_j -p_c -1$ and $(\alpha_y)_r=-c$. In this case,
 $\langle 0|\phi_{\alpha_x}(\pb[r])\phi_{\alpha_y}(\pb[r])|0\rangle \neq 0$.

Similarly, one may say that there exist a part $\alpha_x$ such that $Na^c_i +p_c=[N\alpha_x /r]$ and
$c_N=(N\alpha_x)_r$, and a part $\alpha_y$, with $[N\alpha_y/r]=Nb^c_j -p_c -1$ and $(N\alpha_y)_r=-c_N$. In this case,
 $\langle 0|\phi_{N\alpha_x}(\pb[r])\phi_{N\alpha_y}(\pb[r])|0\rangle \neq 0$.
 \er

One obtains for a pair of $a^c$ and $b^c$ of the same color $c$ (we will denote here $a=ra^c+c$, $b=rb^c+r-c$)
that (\ref{pair-corr}) is equal to
\be\label{sum-hh}
 \langle 0| \phi_{-c} \phi_{c} |0\rangle h_{b^c+1}\{2p_k'\}h_{a^c }\{2p_k'\}+ \cdots +
 \langle 0| \phi_{-a} \phi_{a} |0\rangle h_{a^c + b^c +1}\{2p_k'\}h_{0}\{2p_k'\}
\ee
 $$
 = (-1)^c h_{b^c+1}\{2p_k'\}h_{a^c}\{2p_k'\}+ \cdots +(-1)^{a +1} h_{a^c + b^c }\{2p_k'\}h_{1}\{2p_k'\}
 + (-1)^a h_{a^c + b^c +1}\{2p_k'\}h_{0}\{2p_k'\}
 $$
 For instance, for $r=3$ (i.e. there is the single color $c=1 =\frac 12 (r-1)$), if taking $a^1=a^2=0$,
i.e. $a = 1$, $b=2$, one obtains
 \be
 \langle 0| \phi_2(\pb[3])\phi_1(\pb[3]) |0\rangle =\langle 0| \phi_{-1}\phi_1 |0\rangle h_1\{2p_k'\}h_0\{2p_k'\}
 =(-1)^1  h_1\{2p_k'\}h_0\{2p_k'\} = -2 p_1
 \ee
Now notice that, since $r,j$ are odd, one has from (\ref{h})
 \be\label{h(odd-times)}
 h_i\{2p_k'\}=(-1)^i h_i\{-2p_k'\}
 \ee
therefore (\ref{sum-hh}) is written as
\be\label{sum-hh2}
  (-1)^{c+a^c} h_{b^c+1}\{2p_k'\}h_{a^c}\{-2p_k'\}+ \cdots +
  (-1)^{a} h_{a^c + b^c }\{2p_k'\}h_{1}\{-2p_k'\}
 + (-1)^a h_{a^c + b^c +1}\{2p_k'\}h_{0}\{-2p_k'\}\nn\\
 \ee
 (the parity of $a=ra^c +c$ is equal to that of $a^c+c$ because $r$ is odd).

 We compare (\ref{sum-hh2}) with the one-hook Schur function \cite{Mac}
\be
S_{(j|k)}\{2p_k'\} =
(-1)^{k} \sum_{i=0}^{k} h_{j+i+1}\{2p_k'\}h_{k-i}\{-2p_k'\}
\label{s_hook_bilinear}
\ee
and obtain that
\be
\langle 0|\gamma(\pb[r]) \phi_{rb^c+c^*}\phi_{ra^c+c} |0\rangle =
(-1)^{b^c+ra^c+c} S_{(a^c|b^c)}\{2p_k'\}=
(-1)^{b^c+a^c+c} S_{(a^c|b^c)}\{2p_k'\}
\ee

The case of sub-partition $\mu$ can be considered in a similar way. For a pairwise averages, using (\ref{dynamics})
in the same way as before, and using (\ref{h=q}) and (\ref{Q_ij}), one gets
 \be
\langle 0|\phi_{r\mu}(\pb[r])\phi_{r\nu}([\pb[r]]) |0\rangle
=\langle 0|  \sum_{m\ge 0}\phi_{\mu-mr}h_m\{2p_k'\} \sum_{n\ge 0}\phi_{\nu-nr}h_n\{2p_k'\}  |0\rangle
=2^{-1}Q_{(\mu,\nu)}\{p_k\}
 \ee
 (here $\mu,\nu$ are numbers). The Pfaffian of the Wick theorem yields the projective Schur function labeled by the
 partition $\mu$.

As one can see, after re-numbering, the neutral fermions from complementary groups $a^c$ and $b^c$ are quite similar
to the charged fermions (\ref{psi-pairing}),
while the neutral fermions in the group $\mu$ up to re-numbering the Fourier modes remain to be
neutral (\ref{b-pairing}) inside average (\ref{Q_R}).

Finally, applying the Wick theorem to evaluate averages with all three groups of fermions, one obtains (\ref{Q-S}).

\section{Multicomponent KP and BKP $\tau$-functions \label{malticomponent}}

Following \cite{JM}, one can define two-component charged fermions by $\psi^{(i)}_j=\psi_{2j+i}$,
$\psi^{\dag(i)}_j=\psi^\dag_{2j+i}$, where $i=1,2$ is the component number. Due to (\ref{canonical-charges}),
we see that the fermions with different components anticommute. For each given component, one can construct
the neutral fermions labeled with the same component number via (\ref{phi-psi}). The neutral fermions related to
different components anticomute. We construct currents labeled with the component number and the corresponding
evolution operators,
\be
\gamma^{(i)}(\pb^i)=e^{\sum_{k>0,\,{\rm odd}} \frac 1k p_k^{(i)}J^{(i)}_k  },\quad
J^{(i)}_k=\sum_{n\in\mathbb{Z}} (-1)^n \phi^{(i)}_{-k-n}\phi^{(i)}_n,\quad i=1,2
\ee
Then any vacuum expectation value
\be\label{2-BKP-tau}
\tau(\pb^{(1)},\pb^{(2)})=\langle 0|\gamma^{(i)}(\pb^{(1)})\gamma^{(i)}(\pb^{(2)})e^{\cal A} |0\rangle
\ee
where ${\cal A}$ is quadratic in the neutral fermions is a two-component BKP $\tau$-function, and the sets $\pb^{(1)}$
and $\pb^{(2)}$ are called the sets of 2-BKP higher times.
There is an important
\br\label{BKP-2-BKP}
If one fixes the set $\pb^{(2)}$ (the set $\pb^{(1)}$), then $\tau(\pb^{(1)},\pb^{(2)})$ can be considered as a one-component BKP
$\tau$-function with respect to the set $\pb^{(1)}$ (to the set $\pb^{(2)}$).
\er
Details may be found in \cite{JM} or in \cite{KvdLbispec}.

\end{document}